\documentclass[fleqn,usenatbib]{mnras}

\usepackage{newtxtext,newtxmath}
\DeclareRobustCommand{\VAN}[3]{#2}
\let\VANthebibliography\thebibliography
\def\thebibliography{\DeclareRobustCommand{\VAN}[3]{##3}\VANthebibliography}

\usepackage{lipsum}

\usepackage{graphicx}	% Including figure files
\usepackage{amsmath}	% Advanced maths commands

\usepackage{booktabs} %fancy tables
\newcommand{\knee}{\mathrm{knee}}

\title[Horndeski LSS with LIM]{A Forecast for Large Scale Structure Constraints on Horndeski Gravity with Line Intensity Mapping}

\author[B.R. Scott et al.]{
Bryan R Scott,$^{1}$\thanks{E-mail: bryan.scott@email.ucr.edu}
Kirit~S.~Karkare,$^{2,3}$
Simeon Bird$^{1}$
\\
$^{1}$ Department of Physics $\&$ Astronomy, University of California, Riverside, Riverside, CA 92521, USA\\
$^{2}$ Kavli Institute for Cosmological Physics, University of Chicago, Chicago, IL 60637, USA \\ 
$^{3}$ Fermi National Accelerator Laboratory, MS209, P.O. Box 500, Batavia, Illinois 60510, USA\\
}

\date{Accepted XXX. Received YYY; in original form ZZZ}

\pubyear{2022}

\begin{document}
\label{firstpage}
\pagerange{\pageref{firstpage}--\pageref{lastpage}}
\maketitle

\begin{abstract}
We consider the potential for line intensity mapping (LIM) of the rotational CO(1-0), CO(2-1) and CO(3-2) transitions to detect deviations from General Relativity from $0 < z < 3$ within the framework of a very general class of modified gravity models, called Horndeski theories. 
Our forecast assumes a multi-tracer analysis separately obtaining information from the matter power spectrum and the first two multipoles of the redshift space distortion power spectrum. To achieve $\pm 0.1$ level constraints on the slope of the kinetic gravity braiding and Planck mass evolution parameters, a mm-wave LIM experiment would need to accumulate $\approx 10^8-10^9$ spectrometer hours, feasible with instruments that could be deployed in the 2030s. Such a measurement would constrain large portions of the remaining parameter space available to Scalar-Tensor modified gravity theories. Our modeling code is publicly available.
\end{abstract}


%%%%%%%%%%%%%%%%%%%%%%%%%%%%%%%%%%%%%%%%%%%%%%%%%%

%%%%%%%%%%%%%%%%% BODY OF PAPER %%%%%%%%%%%%%%%%%%

\section{Introduction}

The theory of General Relativity (GR) has withstood attempts at revision on theoretical and experimental grounds for more than a century. In light of the non-renormalizability of GR and the need to explain the observed change in the expansion rate of the universe, there is now a rich taxonomy of theories that revise standard GR, including f(R), Horava-Lifshitz, and scalar-tensor theories \citep[for a thorough review, see][]{Clifton_2012}. Despite stringent experimental limits on deviations from GR on small scales, measurements of the Hubble constant \citep{Riess_1998,Schmidt_1998,Perlmutter_1999}, the Cosmic Microwave Background (CMB; \cite{Planck_2020}) and Baryonic Acoustic Oscillations (BAO; \cite{Wang_2006}) all point to an accelerated expansion of the universe. Although the most minimal explanation is arguably a cosmological constant, other potential solutions include a new coupling to the matter sector or a modification of the gravity theory itself.

Astrophysical and cosmological tests of GR are also worth pursuing even if they do not seek to explain the expansion of the Universe, as they are able to probe regimes inaccessible to Solar System probes. 
Horndeski theories  are the most general theories which include a scalar field and have 2nd order equations of motion \citep{Horndeski_1974}. They are interesting because they include a number of previously studied classes of model as subcases, including Brans-Dicke, f(R), and Galileon models \citep{Bellini_2014}.

Modified gravity effects can be observed  both through changes to the background expansion and large-scale geometry, or through measurements of large-scale structure (LSS) that probe changes to the Poisson equation and the physics of galaxy formation and evolution (\cite{doi:10.1142/11090} and references therein). BAO and CMB measurements probe geometry, while lensing, Redshift Space Distortions (RSD), and biased tracers of the matter power spectrum probe structure formation. It is typical to combine multiple probes to improve constraining power and perform consistency checks \citep{Troxel2018}, with specific consistency conditions for Horndeski theories derived in \cite{Hojjati2011, P2018}. Measurements of galaxy cluster abundance and the linear growth rate of perturbations have placed limits on modifications to gravity, especially in the dark energy equation of state ($w$) - linear growth rate parameter ($\gamma$) plane (cf. \cite{2008MNRAS.387.1179M, Rap2008}).

As a general framework of modified gravity, Horndeski theories have a variety of operators which can be constrained by different cosmological experiments.
In the context of effective field theory parameterizations of modified gravity, \cite{Kreisch_2018} find that modified gravity constraints are primarily driven by a large amplitude modification of the ISW in CMB measurements. \cite{Spurio_Mancini_2019} pursue lensing and clustering constraints on Horndeski model parameters from KiDS + GAMA. \cite{Noller_2019} combine measurements of the CMB, RSD and SDSS $P_{\mathrm{m}}(k)$ measurements from the LRG catalog to derive constraints on the Horndeski model parameters, with results similarly driven by the large effect of modified gravity parameters at low $\ell$ in  Planck $C_{\ell}^{\rm TT}$ measurements. However, large uncertainties, especially from galaxy biasing and degeneracies in the measurement of $P_{\mathrm{m}}(k)$, limit its constraining power beyond the CMB-only result. In addition, the recent simultaneous detection of gravitational and electromagnetic waves from neutron star mergers constrains the speed of gravity to match the speed of light to high accuracy \citep{2017PhRvL.119y1301B, 2018PhRvD..97j4038A}. Interactions between dark energy and gravitational waves can also be used to constrain a linear combination of Horndeski model parameters \citep{2020JCAP...05..002C, 2020PhRvD.101f3524N}. Future theoretical work as well as next generation-surveys (e.g., Vera C. Rubin Observatory/Legacy Survey of Space and Time \citep{2019ApJ...873..111I}, Nancy Grace Roman Space Telescope \citep{2015arXiv150303757S}, Euclid \citep{2011arXiv1110.3193L}) are likely to improve measurements of the matter power spectrum, and will improve constraints subject to uncertainties in the biasing of LSS tracers.

In this paper we will make forecasts for an alternative avenue for constraining modified gravity, based on Line Intensity Mapping (LIM). 
LIM is a promising technique for constraining both astrophysics and cosmology in large cosmological volumes \citep{Kovetz_2017, Karkare_2018, Creque2018, Dizgah2019, Gong2020, Pullen_2014, Breysse2014}.
LIM uses moderate resolution observations to detect LSS in aggregate by integrating over discrete sources \citep{Kovetz_2017}. By using a spectrometer to target a spectral line with known rest-frame wavelength, the redshift of the source emission is known and 3D maps of cosmic structure can be constructed.
LIM is particularly advantageous for measuring a wide redshift range, including high redshifts where individual galaxies become too faint to be detected in a traditional galaxy survey. This makes LIM an attractive way to probe the evolution of structure during both the epochs of matter and dark energy domination, and thus provide powerful constraints on modified gravity scenarios.

Numerous atomic and molecular lines are now being targeted by LIM experiments across a wide range of wavelengths. High-redshift measurements of the 21 cm neutral hydrogen spin-flip transition in the intergalactic medium are expected to probe the timing and sources of reionization \citep[e.g.~HERA][]{DeBoer_2017}, while similar measurements of HI within galaxies at lower redshifts will constrain dark energy \citep{wu2021prospects, cosmicvisions21cmcollaboration2019inflation}. At shorter wavelengths, far-IR lines such as the CO $J\rightarrow J-1$ rotational transitions and the ionized-carbon [CII] fine structure line trace star formation and cold gas, are known to be bright in early galaxies, and are being targeted at redshifts $0 < z < 10$ by instruments observing in the cm-THz range \citep{Righi_2008, Breysse2014, Li2016, Fonseca2016, Padmanahhan2017, Yue2015, silva2015prospects}.  Finally, lines such as H$\alpha$ and Ly$\alpha$ are targets for optical LIM experiments; for example, SPHEREx targets galactic astrophysics and reionization \citep{Gong2011, Silva2013, silva2017tomographic, 2017ApJ...835..273G}. 

Current and near-future far-IR LIM experiments aim to constrain the star formation rate and galaxy formation history over a wide range of redshifts. Those targeting CO include COPSS \citep{2016ApJ...830...34K} and mmIME \citep{Keating2020, breysse2021estimating}, which have both reported first detections of CO shot noise power, COMAP \citep{Li2016}, EXCLAIM \citep{padmanabhan2021intensity}, YTLA \citep{YT2009} and SPT-SLIM \citep{Karkare_2022}. TIME \citep{2014SPIE.9153E..1WC} and CONCERTO target [CII] \citep{Ade2020}. Recently the COMAP collaboration has placed upper limits on the CO(1-0) signal at clustering scales \citep{Havard_2021}.  While the first generation of mm-wave LIM experiments will primarily constrain the astrophysics of the emission line, future generations have the potential to deliver competitive constraints on cosmology, including early/dynamical dark energy and neutrino masses \citep{Karkare_2018, dizgah2021neutrino}. In particular, compact mm-wave spectrometers are now being demonstrated that could enable future surveys with orders of magnitude more sensitivity than current experiments \citep{Karkare2020}.

In this paper, we investigate the constraining power of future ground-based wide-bandwidth mm-wave LIM experiments targeting multiple rotational CO transitions over the redshift range $z \approx 0-3$. In addition to large accessible cosmological volumes, this extends constraints on modified gravity to higher redshifts than are available in current large optical surveys. In section \ref{sec:horndeski} we review Horndeski gravity and the application to LSS through the matter power spectrum and redshift space distortions. Then, in section \ref{sec:lim}, we introduce the formalism of LSS measurements with LIM. In section \ref{sec:results}, we investigate the range of accessible scales and required survey integration times to achieve competitive constraints on the linear theory parameters, accounting for the atmosphere, astrophysical continuum, and interloper lines. In section \ref{sec:discussion}, we discuss implications of these results. We conclude in section \ref{sec:conclusion}. 

Although modifications to GR generally imply a different expansion history, we assume that all deviations are small and only affect linear structure formation around a $\Lambda$CDM background. As such, where necessary, we assume a flat $\Lambda$CDM-like cosmology with $h = 0.678$, $\Omega_b h^2 = 0.0224$, $\Omega_c h^2 = 0.12$, and $\Omega_\Lambda$ = $1 - \Sigma \Omega_i$. 

\section{Horndeski Gravity}
\label{sec:horndeski}

Horndeski theories construct a relativistic theory of gravity from a Lagrangian including a metric tensor and a scalar field, and lead to second order equations of motion \citep{Horndeski_1974}. In this section, we review the features of Horndeski theory relevant to this work, and refer the interested reader to \cite{Bellini_2014}, which develops the formalism employed here, and its application to the Einstein-Boltzmann solver \texttt{CLASS} \citep{Blas_2011} to produce the extended Horndeski in Linear Cosmic Anisotropy Solving System (\texttt{HI\_CLASS}) \citep{Zumalacarregui_2017, Bellini_2020}. 

In the linear regime, solving the perturbed Einstein equation allows for the construction of four functions of time, denoted $\alpha_i(t)$, that translate the functional degrees of freedom in the action into four time-dependent parametric degrees of freedom \citep{Bellini_2014}. The Horndeski action, the background relations, and prescriptions for the $\alpha_i(t)$ fully determine the evolution of perturbations in the linear regime and hence LSS. There are $4$ functions, two of which are in principle measurable by LIM ($\alpha_B$ and $\alpha_M$) and two of which are not ($\alpha_K$ and $\alpha_T$). They have the following physical interpretations:
\begin{itemize}
    \item $\alpha_B$ encodes mixing between the scalar and metric perturbations that arises from the clustering of the Horndeski scalar field, and appears as perturbations to $T_{0i}$. $\alpha_B = 0$ in $\Lambda$CDM + GR. We treat $\alpha_B$ as a free parameter to be constrained by the LIM experiment. 
    \item $\alpha_M$ rescales the Planck mass, representing a change in the strength of gravity. While a constant rescaling of the strength of gravity does not affect structure formation, its time evolution generates anisotropic stress. Since $\alpha_M$ parameterizes the evolution of the Planck mass with time, $\alpha_M = 0$ in $\Lambda$CDM + GR. We treat $\alpha_M$ as a free parameter to be constrained by the LIM experiment.

    \item $\alpha_K$ represents perturbations to the energy-momentum tensor $T_{\mu\nu}$ arising directly from the action. These can be thought of as perturbations in an additional fluid connected with the modification to gravity. However, $\alpha_K$ affects only scales close to the cosmological horizon, far larger than those measured by LIM or other LSS probes (for a discussion of this to second order, see \cite{Bellini_2014}). While $\alpha_K = 0$ represents the value in $\Lambda$CDM + GR and is therefore a natural choice, we choose $\alpha_K = 1$ to ensure that our models easily satisfy the condition for avoiding ghosts in the scalar mode: $\alpha_K + 3/2\alpha_B^{2} > 0$.

    \item $\alpha_T$ gives the tensor speed excess, potentially inducing anisotropic stress, even in the absence of scalar perturbations. $\alpha_T = 0$ in $\Lambda$CDM + GR. We set $\alpha_T = 0$, as it is well constrained by measurements of the speed of gravitational waves \citep{2017PhRvL.119y1301B, 2018PhRvD..97j4038A}.
  \end{itemize}

These expressions are implemented in the Einstein-Boltzmann solver Horndeski in Cosmic Linear Anisotropy Solver (\texttt{HI\_CLASS}), which we use to predict the matter power spectrum $P_{\mathrm{m}}(k)$ under an assumed $\Lambda$CDM background, and to vary the free functions according to the parameterizations described in the next section. 

\subsection{Parameterizations}
\label{sec:parameterizations}

In the linear regime of cosmological perturbation theory, we assume that all perturbations are small and taken around a flat background spacetime,
\begin{equation}
    ds^2 = -(1+2\Psi)dt^2 + (1+2\Phi)\delta_{ij}dx^i dx^j,
\end{equation}
where $\Psi$ and $\Phi$ are small metric perturbations. In the case of fluid scalar perturbations and general theories of gravity, LSS observations such as the galaxy power spectrum, weak lensing shear field, or RSD probe a small number of combinations of these potentials. In Horndeski theory the potentials are also complicated functions of the $\alpha_i$, arbitrary functions that represent the maximal amount of information available from cosmology to constrain the dynamics of this class of models. The evolution of the flat background itself can be determined from the Friedmann equations.

The functional freedom to pick the Horndeski $\alpha_i$ allows any evolution for the background spacetime to be realized. LSS alone cannot pick out either the expansion history or a unique form for the $\alpha_i$. To reduce this freedom, we begin by first noting that geometric measurements are consistent with the universe being nearly $\Lambda$CDM, which we select as our model for the background evolution. Once a background is chosen, it is necessary to define a functional form to parameterize how modifications to gravity evolve with time. 

\textbf{Parameterization I}: A natural choice in a nearly-$\Lambda$CDM universe is to parameterize the modified gravity effect as proportional to the cosmological constant density, $\Omega_{\Lambda}$. As this term grows with redshift, it ``turns on'' modified gravity effects at late times and during the epoch of dark energy domination. Thus for our first parameterization we assume that $\alpha_B$ and $\alpha_M$ are linear functions of $\Omega_{\Lambda}$: 
\begin{equation}
    \alpha_{B,M} = c_{B,M} \Omega_{\Lambda}.
\end{equation}
Here we have adopted notation from \cite{Noller_2019} and \cite{Kreisch_2018}, and refer to this as Parameterization I. 

\textbf{Parameterization II}: 
To evaluate the sensitivity of our probes to observations at high redshift, we use an alternate parametrization where the effect of gravity modification is linearly proportional to the scale factor \citep{Zumalacarregui_2017}. This allows the modified gravity to become important at early times, before the onset of dark energy domination. We thus have
\begin{equation}
    \alpha_{B,M} = c_{B,M} a. 
\end{equation}

\begin{figure*}
\centering
\includegraphics[scale = 0.22]{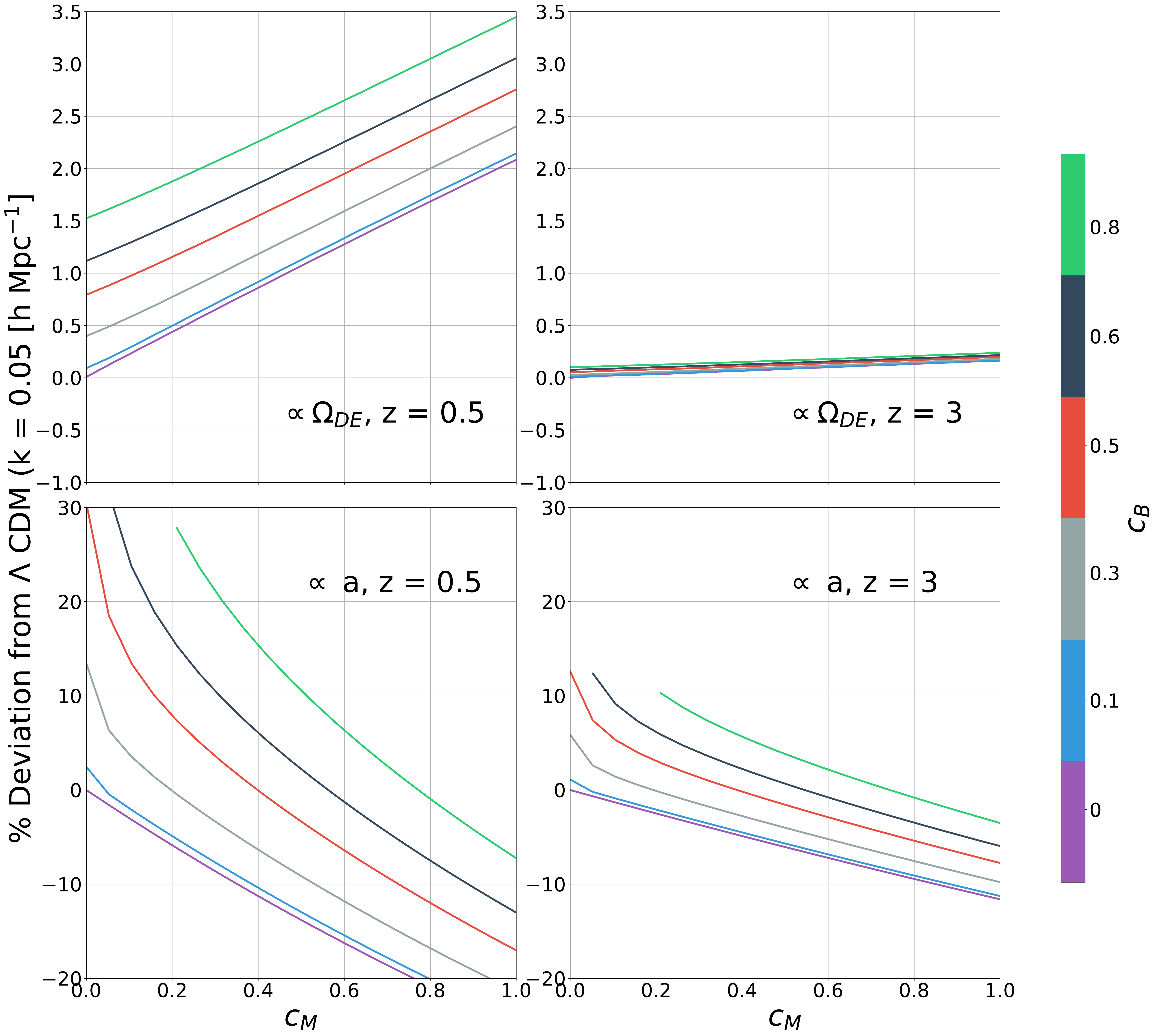}
\caption{\label{fig:Pk_dev}Relative deviation of the matter power spectrum for fixed $k=0.05$ h Mpc$^{-1}$ at $z=0.5$ (Left) and $z=3$ (Right) as a function of $c_M$, with curves labelled by their value of $c_B$. Top row shows Parametrization I, bottom row shows Parametrization II. 
The $c_B$ and $c_M$ parameters are allowed to vary over the range $0-1$. In Parametrization II ($\alpha_i \propto a$) we have truncated the results due to gradient instabilities when $c_M$ is small and $c_B$ is large.}
\end{figure*}

Figure~\ref{fig:Pk_dev} shows the relative deviation of the matter power spectrum at a fixed scale ($k=0.05$ h Mpc$^{-1}$) as the Planck mass rescaling $c_M$ and braiding $c_B$ parameters are allowed to vary. Large deviations from $\Lambda$CDM are possible for extreme values of the $\alpha$ functions. As noted in \cite{Noller_2019}, curves that intersect the $\Lambda$CDM prediction exhibit a degeneracy between $c_M$ and $c_B$ for the matter power spectrum with $c_B \approx 1.8 c_M$ in Parameterization II ($\alpha_i \propto a$). By design, the effect of modifying gravity is largest for the late-time universe, near the end of matter domination. One implication of this evolution is that achieving robust constraints on these linear theory parameters and simultaneously constraining deviations from GR in both parameterizations requires an experiment that targets a large range in redshift. 

By selecting a fiducial $k$-scale to summarize the effects of varying $\alpha_M$ and $\alpha_B$, we have ignored the $k$-dependence introduced by the modification to gravity. A well known generic feature of these models is a turnover in the power spectrum, where an excess on large scales becomes a deficit on small scales (with respect to $\Lambda$CDM). However, this turnover occurs on scales near the cosmological horizon and is thus extremely difficult to measure with LSS measurements.\footnote{The turnover is an unambiguous signature of modified gravity, and would constrain the braiding scale, a function of $\alpha_M$ and $\alpha_B$.} 

At the intermediate scales measured by a LIM experiment, the characteristic feature of modified gravity models relative to $\Lambda$CDM + GR is a uniform excess in the power spectrum. The size and behavior with varying $c_M, c_B$ of the effect depends strongly on the choice of parameterization, with a $<1\%$ difference in $\Lambda$CDM at $z=3$ in Parameterization I and a few percent difference at $z=0.5$, even for extreme values of the $c_M, c_B$. The effect of modified gravity on the power spectrum is larger in Parameterization II, approaching instability in the theory when $c_M$ is small and $c_B$ is large. We therefore expect greater sensitivity to the $c_M, c_B$ in Parameterization II than in Parameterization I.

To summarize, we have two modified gravity functions that scale with the background evolution of the spacetime that we seek to constrain. The $c_B$ and $c_M$ parameters govern respectively the evolution of the braiding $\alpha_B$ (clustering of dark energy) and the Planck mass run rate $\alpha_M$ (the large-scale strength of gravity). $\Lambda$CDM differs from the models we consider in that the large-scale strength of gravity is fixed and dark energy does not cluster. We assume two functional forms for the scaling of these parameters with the background evolution: one parameterization that scales with the effective dark energy component density $\Omega_{DE}$ and one that scales with the scale factor. We fix the remaining two functions to be constant and assign them to unity. We specifically forecast for the uncertainties $\sigma(c_B)$, $\sigma(c_M)$.

\begin{figure*}
\centering
\includegraphics[scale=0.2]{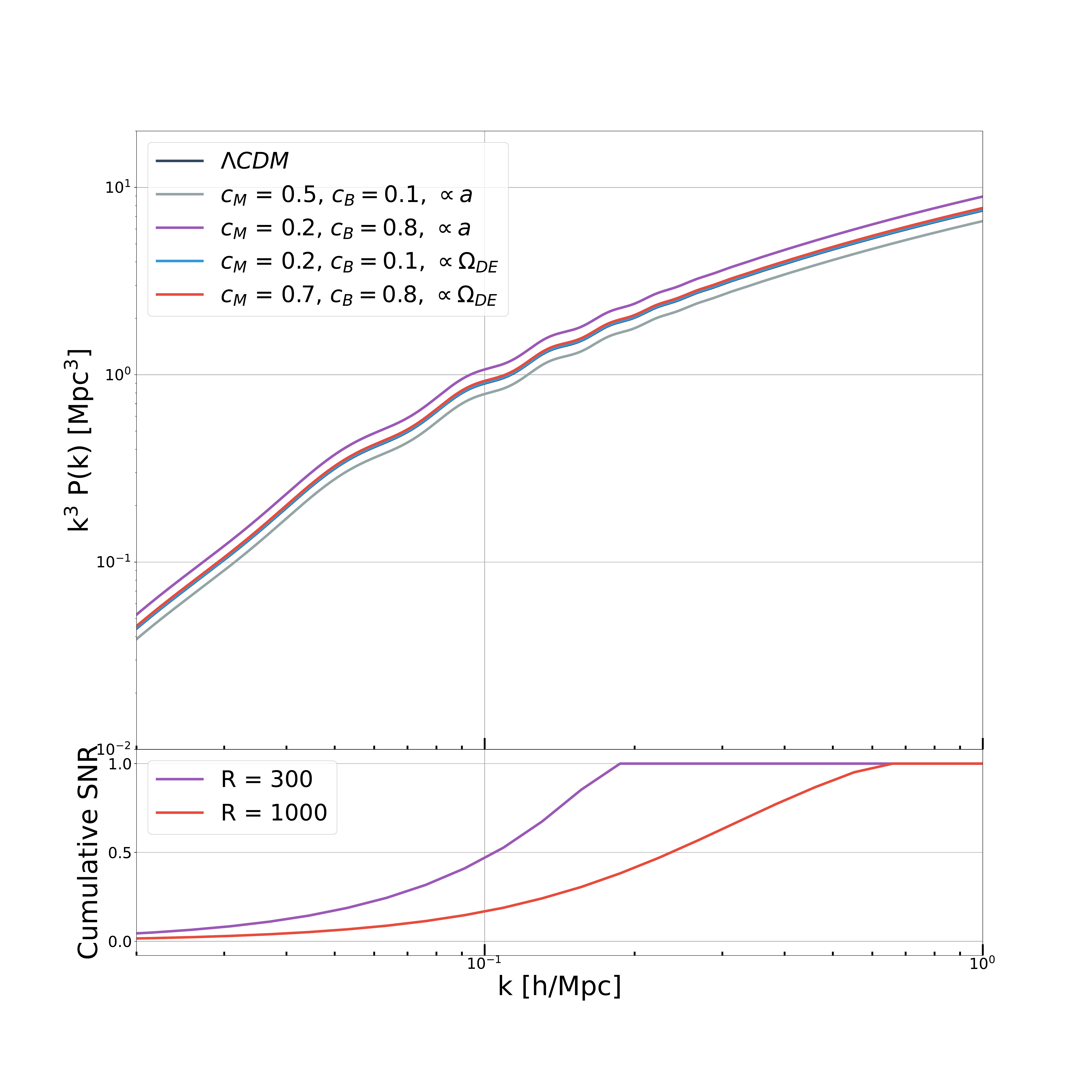}
\caption{\label{fig: relevant_scales} Top panel shows the matter power spectrum at $z=0.5$ in both parameterizations. We have chosen values of $c_B$ and $c_M$ representative of the range of deviations in $P_m(k)$ that we constrain. Bottom panel shows the cumulative constraining power as a function of scale assuming a spectral resolution of $R=300$ or $1000$ in the baseline $f_{\mathrm{sky}}=40\%$ case. The SNR saturates once the scales probed are below the spectral resolution of the LIM experiment. The lower spectral resolution with $R=300$ causes the SNR to saturate at an larger $k$ than in the $R=1000$ case.}

\end{figure*}

\begin{figure*}
\centering
\includegraphics[scale=0.23]{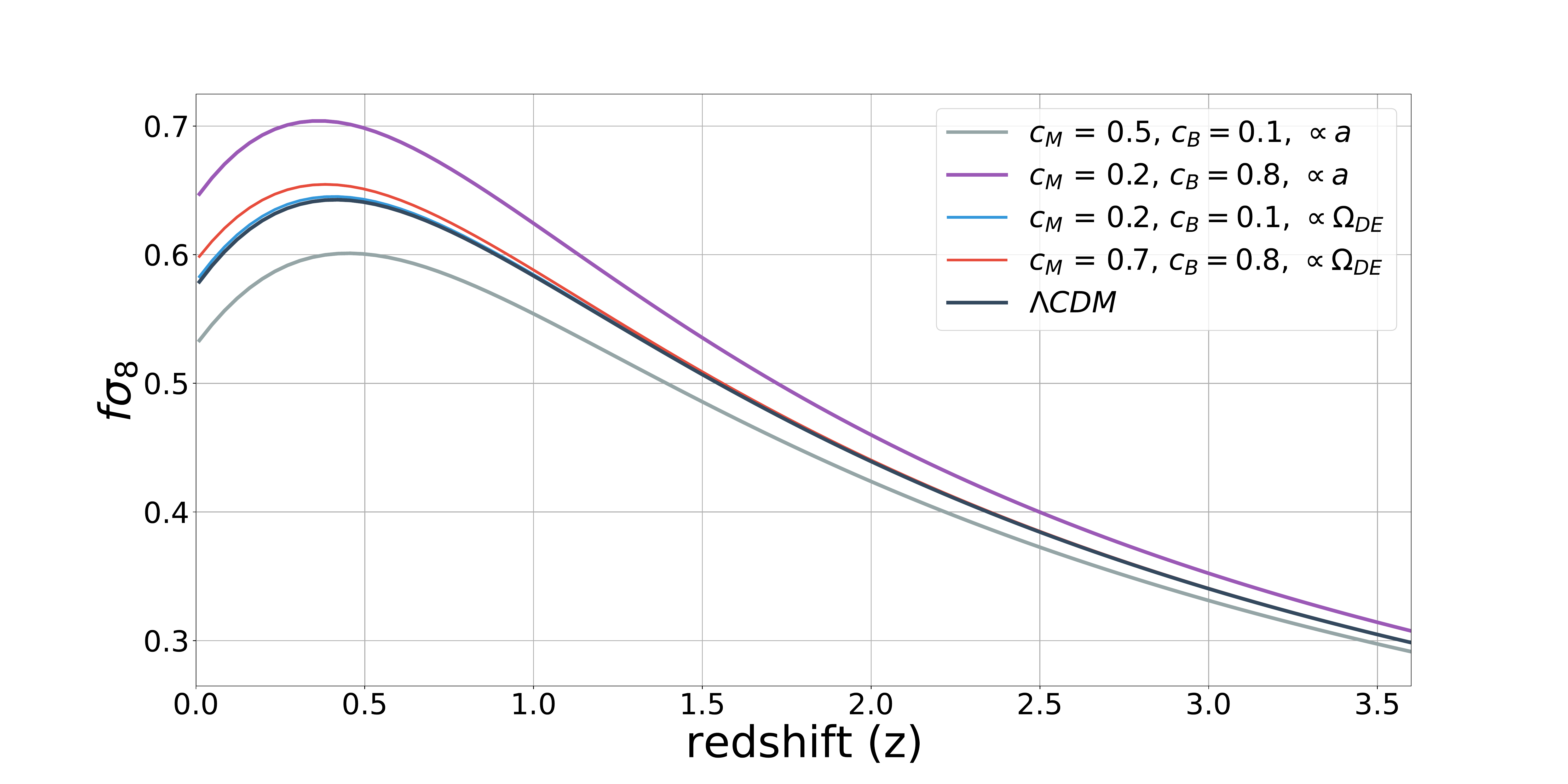}
\caption{\label{fig:Pk}  Evolution of the power spectrum normalization $f(z) \sigma_8(z)$ over the range of redshifts accessible to the experiment we forecast for. We have also indicated the approximate redshift where the CO(1-0), CO(2-1), and CO(3-2) lines that we target in our forecast are brightest. The evolution of the line brightness differs from that of the power spectrum normalization, both in the peak and evolution with redshift. This allows an experiment that measures LSS at a range of redshifts to disentangle the evolution of a modified gravity effect on $P_{\mathrm{m}}(k)$ from that of $I(z)$.}
\end{figure*}

\section{Line Intensity Mapping}
\label{sec:lim}

In this section we discuss the details of the LIM observables used in our projections. We then discuss experimental effects that limit the scales accessible in the power spectrum, in addition to the effects of interloper lines and Galactic foregrounds.

\subsection{Line Power Spectrum}
\label{sec:models}

Emission lines targeted by LIM experiments originate in galaxies that are biased with respect to the underlying matter overdensity field. On the intermediate and large scales we consider here, outside the nonlinear regime, we can parameterize clustering with a scale-independent clustering bias $b(z)$ that varies with redshift. Since the target lines we consider are correlated with galaxy properties (e.g., star formation rate and metallicity) that evolve with redshift, the line intensity $I(z)$ is also redshift-dependent. The LIM clustering power spectrum is
\begin{equation}
    \label{eqn:Pclust}
    P_{\mathrm{clust}}(k,z) = b^2(z)I^2(z)P_{\mathrm{m}}(k,z).
\end{equation}
Here $P_{\mathrm{m}}(k,z)$ is the underlying matter power spectrum which contains the cosmological information (Section~\ref{sec:parameterizations}). We show the matter power spectrum for a range of choices of the $c_M$ and $c_B$ in Figure~\ref{fig: relevant_scales},
and the product of the matter power spectrum normalization and linear growth function, $f(z)\sigma_8(z)$, in Figure~\ref{fig:Pk}. We assume the bias is scale independent and varies linearly with redshift, b = (1 + z), which is a reasonably close approximation to the Sheth-Torman-inspired \citep{2001MNRAS.323....1S} bias evolution from \cite{dizgah2021neutrino}.

We assume that the line evolution is given by the line models in \cite{delabrouille2019}. Line intensities are estimated from the specific luminosity density $\rho(z)$. Redshift-dependent luminosity densities are a function of the halo mass ($\mathrm{dn/dM(M,z)}$) or line luminosity functions ($\mathrm{dn/dL(z)}$), that are obtained through empirical scaling relations through the dependence of L(M) with SFR(M,z) or SFRD(M,z). In the \cite{delabrouille2019} models, the line luminosities are derived from the Eagle simulation and are uncertain from a factor of a few to ten. For a recent review of line intensity modeling, see \cite{Bernal_2022}. %\ksk{Fix math italics here}

The observable effect of the modification to gravity is a constant excess or deficit in $P_{\mathrm{m}}(k)$. The degeneracy between the line intensity, $I(z)$, and the matter power spectrum introduces a degeneracy between the modified gravity effect and the evolution of the bias and line intensity. However, $f(z)\sigma_8(z)$, which controls the power excess on the scales we observe, has a distinctly different form of redshift evolution from the astrophysics-dependent term $I(z)$. Observing a continuing increase relative to the $\Lambda$CDM expectation as the line intensity decreases (or vice-versa) as a function of redshift can therefore break this degeneracy, allowing us to potentially recover a signal from modified gravity effects.

A LIM experiment measures the clustering power, shot noise due to the discrete nature of the emitting galaxies, and instrumental noise:
\begin{equation}
    P_{\mathrm{obs}}(k,z) = P_{\mathrm{clust}}(k,z) + P_{\mathrm{shot}}(z) + P_{\mathrm{N}}. 
\end{equation}

The uncertainty in the power spectrum measurement from a LIM experiment depends both on the number of observed modes and on the instrumental noise.  We write the number of Fourier modes at a scale $k$, in bins of width $\Delta k$, in a total volume $V_s$ as
\begin{equation}
\label{Eqn: Nmodes} 
    N_m(k) = \frac{k^2 \Delta k V_s}{4 \pi^2}
\end{equation}
and the variance $\sigma(k)$ on a measurement of $P(k)$ at a scale $k$ is 
\begin{equation}
    \label{Eqn: variance}
    \sigma^2(k,z) = \frac{P^2_{\mathrm{obs}}(k,z)}{N_m(k)}. 
\end{equation}

Estimates for $P_{\mathrm{shot}}$ are given in Table \ref{tab:targets}, while estimates of $P_\mathrm{N}$ are discussed in Section \ref{sec:noise}.

\subsection{Redshift Space Distortions}
\label{sec: RSD}

Observations of LSS are not made in the isotropic comoving space in which the matter power spectrum is defined, but in the 2+1 dimensional space of angles and redshift. Since the redshift of an emitter has components due to both the Hubble flow and its peculiar velocity, the power spectrum in redshift space is distorted relative to comoving space \citep{Hamilton_1998}.

Because the inferred transverse and line of sight coordinates are affected differently by the RSD, it is necessary to consider the full anisotropic power spectrum in the space of parallel ($k_{||}$) and perpendicular ($k_{\perp}$) modes. The RSD power spectrum can be expressed in ($k, \mu$) coordinates, where the cosine of the angle is denoted $\mu = \hat{z} \cdot \hat{k}$. Then, in the linear plane-parallel approximation, the anisotropic matter power spectrum is
\begin{equation}
    \label{eqn: RSD_basic} 
    P_{\mathrm{obs}}(k,\mu, z) = [b(z)^2 I(z)^2 + f(z)^2 I(z)^2 \mu^2]^2 P_{\mathrm{m}}(k).
\end{equation}

Here $f(z)$ is the linear growth rate of structure, which is sensitive to modifications of gravity. We show the redshift evolution of $f \sigma_8$ for the range of Horndeski theories we consider in Figure \ref{fig:Pk}. 

\citet{Noller_2019} consider constraints on the Horndeski theory parameters $c_B$ and $c_M$ from both anisotropic clustering measurements of the growth factor $D$ in SDSS and $f\sigma_8$ at $z=0.57$ in the 6dF survey at $z=0.067$. While the power spectrum adds little constraining power directly, the RSD constraint improves the posterior uncertainties, especially on the $c_M$ parameter, when compared to the CMB-only constraint.

Although one can infer the value of $f\sigma_{8}$ directly from the full shape of the anistropic matter power spectrum in $(k, \mu, z)$ space, it is simpler to consider constraints from the non-vanishing $\ell=0,2, 4$ moments obtained by convolving Eq.~\ref{eqn: RSD_basic} with the Legendre polynomials $\mathcal{L}_l$:
\begin{equation}
    \label{eqn: Legendre} 
    P_{l}(k) = \frac{2l+1}{2} \int_{-1}^{1} \mathcal{L}_l(\mu) P(k, \mu) \ d\mu
\end{equation}

Explicit expressions for the monopole and quadrupole are
\begin{align}
    \label{eqn: non-vanishing}
P_0(k) = \left(1+\frac{2}{3} \beta (bI)^2 + \frac{1}{5}  (bI)^2 \beta^2 \right)P_{\mathrm{m}}(k) \nonumber\\
P_2(k) = \left(\frac{4}{3} (bI)^2 \beta + \frac{4}{7}(bI)^2 \beta^2 \right) P_{\mathrm{m}}(k).
\end{align}

Here we neglect the $l=4$ moment since the hexadecapole is both difficult to measure and contains little information not present in the first two multipoles \citep{Chung2019}. For consistency with the literature, we also work with $\beta = f/b$ rather than $f \sigma_8$ directly. These expressions allow us to compute the moments of the redshift space distortions from the isotropic power spectrum $P_{\mathrm{m}}(k)$. The variance between the multipole moments can be computed explicitly:

\begin{equation}
\label{eqn:multi_cov}
    \mathrm{Cov}_{l, l'}(k) = \frac{(2l + 1)(2l' + 1)}{N_{m}}\int_{-1}^{1} \mathscr{L}_l(\mu)\mathscr{L}_{l'}(\mu) (P_{\mathrm{obs}}(k,\mu))^2 \ d\mu. 
\end{equation}

\citet{Taruya_2010} (eq. C2-C4) gives explicit expressions for $\mathrm{Cov}_{l,l'}$ (where we have here combined their shot noise term with our $P_\mathrm{N}$ notation).
For the monopole: 
\begin{multline}
\label{Cl00} 
\mathrm{Cov}_{0, 0}(k) = \frac{2}{N_k} \bigg[\left(1 + \frac{4}{3}\beta + \frac{6}{5}\beta^2 + \frac{4}{7}\beta^3 + \frac{1}{9}\beta^4 \right)  \\ \times
(bI)^2 P_{\mathrm{m}}(k)^2 + 2 P_{\mathrm{N}} \left(1 + \frac{2}{3}\beta + \frac{1}{5} \beta^2 \right) (bI)^2 P_{\mathrm{m}}(k) + P_{\mathrm{N}}^2 \bigg]\,.
\end{multline}
For the monopole-quadrupole cross-term:
\begin{multline}
\label{Cl02} 
\mathrm{Cov}_{0, 2}(k) = \frac{2}{N_k} \bigg[\left(\frac{8}{3}\beta + \frac{24}{7}\beta^2 + \frac{40}{21}\beta^3 + \frac{40}{99}\beta^4 \right) \\ \times 
(bI)^2 P_{\mathrm{m}}(k)^2 + 2 P_{\mathrm{N}} \left( \frac{4}{3}\beta + \frac{4}{7} \beta^2 \right) (bI)^2 P_{\mathrm{m}}(k) \bigg].
\end{multline}
Finally, for the quadrupole:
\begin{multline}
\label{Cl22} 
\mathrm{Cov}_{2, 2}(k) = \frac{2}{N_k} \bigg[ \\ \left(5 + \frac{220}{21}\beta + \frac{90}{7}\beta^2 + \frac{1700}{231}\beta^3 + \frac{2075}{1287}\beta^4 \right) \times 
(bI)^2 P_{\mathrm{m}}(k)^2 \\+ 2 P_{\mathrm{N}} \left(5 + \frac{220}{21}\beta + \frac{30}{7} \beta^2 \right) (bI)^2 P_{\mathrm{m}}(k) + 5 P_{\mathrm{N}}^2 \bigg].
\end{multline}

The above expressions are exact in the case of flat $P_{\mathrm{N}}$, and are approximately correct on intermediate and large scales where the finite spatial and spectral resolution induce only small attenuation in the signal.

\subsection{Target Lines, Redshifts, and Noise Estimates} 
\label{sec:noise}

Our forecasts focus on measuring the LIM power spectrum from $0 < z < 3$, where ground-based CO experiments are most sensitive. 

In Parametrization I ($\alpha_i \propto \Omega_{\Lambda}$), the excess near the turnover in the matter power spectrum at $k = 0.01$ is $\sim 1\%$ at $z=3$, and an order of magnitude larger at $z=0.5$. The evolution in the effect of modified gravity for Parametrization II ($\alpha \propto a$) is comparable in magnitude, but begins at earlier redshifts, as shown in Figure~\ref{fig:Pk_dev}.

We consider an experiment measuring the CO $\rm J \rightarrow \rm J-1$ rotational transitions, which emit at rest-frame frequencies of $115 \rm J$ GHz. CO offers several advantages compared to other LIM targets: it is a known tracer of molecular gas and is therefore indicative of star formation (which peaked at $z \sim 2$), it has been detected in individual galaxies at high redshift using ground-based telescopes observing in the millimeter band, and the multiple transitions allow a wide range of redshifts to be detected in a modest instrumental bandwidth. Our forecasts use the CO line amplitudes from \cite{delabrouille2019}.

To detect the CO power spectrum we consider ground-based mm-wave LIM surveys observing roughly from 75--310 GHz. Technology for this frequency range has seen significant recent development for large-format CMB arrays: focal planes featuring dense arrays of background-limited detectors are now common \citep{bicep_dets}, and current instruments have demonstrated wideband optics that can measure the 1--3 mm band in a single receiver \citep{nadolski2019}.  Current-generation mm-wave spectrometers are significantly larger than their broadband counterparts since they generally use a physically large apparatus (e.g. grating, Fourier Transform, or Fabry-Perot) for spectral separation.  However, \textit{on-chip} spectrometer technology is rapidly progressing \citep{shirokoff2012} and instruments are now being planned to demonstrate LIM with dense spectrometer arrays that approach CMB packing efficiency.  Our forecasts anticipate that this technology can be scaled over the next ten years in the same manner as CMB instruments leading up to CMB-S4 \citep{Abazajian2016}.

The oxygen line at 118 GHz and the water line at 183 GHz naturally divide up the 75--310 GHz mm-wave band into three windows: 75--115 GHz, 120--175 GHz, and 190--310 GHz. We discuss our approach to estimating noise power in Appendix~\ref{app:noise_power}. To account for the frequency dependence of the line temperature, we calculate an effective redshift and line strength for each target by averaging over the window. Target line frequencies, temperatures, and redshifts are given in Table~\ref{tab:targets}. Additional contributions to the noise model are discussed in the following section.

\begin{table}
\caption{Line frequencies, target redshifts, $P_{\mathrm{shot}}$ estimates, and line temperatures used in this forecast. Unlisted rotational transitions up to CO(9-8) are assumed to contribute interloper power, but are not included as targets as they are an order of magnitude smaller in line brightness temperature.}
\label{tab:targets}
\begin{tabular}{@{}llllll@{}}
\toprule
Line    & $\nu_{\mathrm{rest}}$ [GHz] & $\nu_{\mathrm{obs}}$ [GHz] & $z$    & $P_{\mathrm{shot}}$ [$\mu$K$^2$] & Temp [$\mu$K] \\ \midrule
CO(1-0) & 115         & 95                 & 0.21 & 2.66                 &  0.14           \\
CO(2-1) & 230         & 95                 & 1.42 & 8.54                 &  0.75           \\
CO(3-2) & 345         & 95                 & 2.63 & 2.98                 &  0.60            \\
CO(2-1) & 230         & 150                & 0.53 & 81.8                 &  0.24            \\ 
CO(3-2) & 345         & 150                & 1.3  & 100                  &  0.46            \\
CO(3-2) & 345         & 245                & 0.4  & 295                  &  0.14            \\ 
\bottomrule
\end{tabular}
\end{table}

\subsection{Finite Resolution and Foregrounds} 

\subsubsection{Instrument Resolution}

The scales accessible to a LIM experiment are determined by the finite spatial and frequency resolutions. In the frequency direction, the smoothing scale is characterized by the spectrometer resolution $\delta \nu$, while in the transverse direction, the smoothing is a function of the beamwidth $\theta_b$. These correspond to comoving smoothing scales at redshift $z$ in the transverse $\sigma_\perp$ and parallel (to the line of sight) direction $\sigma_{||}$. In the perpendicular (spatial) direction, the smoothing scale is
\begin{equation}
    \sigma_\perp = \frac{\theta_b R(z)}{\sqrt{8 ln 2}},
\end{equation}
where $\theta_b$ is the full width at half maximum of the beam. The smoothing scale in the parallel (frequency) direction is a function of the target frequency, resolution, and the Hubble scale $H(z)$,  
\begin{equation}
    \sigma_{||} = \frac{c \delta\nu (1+z)}{H(z)\nu_{\mathrm{obs}}}.
\end{equation}

The noise power spectrum $P_{\mathrm{N}}(k)$ is the product of the white noise level $P_{\mathrm{N}}$ and a factor accounting for the finite spectral and spatial resolutions of the instrument, 
\begin{equation}
\label{smoothing} 
    P_{\mathrm{N}}(k) = P_{\mathrm{N}} e^{k^2 \sigma_\perp^2} \int_0^1 d\mu \ e^{\mu^2 k^2 (\sigma_{||}^2 - \sigma_\perp^2)}
\end{equation}
where $\mu$ is the cosine of the angle between the wavevector $k$ and the line-of-sight direction, and the integral averages over all such angles $\mu$ to yield the spherically-averaged 3D power spectrum. Here we treat the signal $P(k)$ as fixed, while the finite resolution of the survey causes the noise to become inflated at small scales. This differs from the physical situation, in which instruments generally have flat noise properties as a function of $k$ (or its angular counterpart, $\mathscr{l}$), above a scale $\mathscr{l}_{\mathrm{knee}}$. In fact, it is the inherent signal that is attenuated and not the noise.

Each point in the 2D $k_{||}-k_{\perp}$ Fourier plane (averaged over the angular directions) has some attenuation factor due to the instrument resolution. The RSD introduce some phase dependence into the signal as structures move in the redshift direction only, that are picked out by the RSD operator. However, the attenuated 2D noise does not change due to the velocity-induced distortion of the signal. On large and intermediate scales, and for small values of $P_{\mathrm{N}}$ the attenuation factor contributes negligibly to the noise on measurements of the multipole moments. This allows us to use Eqs.~\ref{Cl00} - \ref{Cl22} even in the case of finite instrument resolution.

\subsubsection{Atmospheric Fluctuations}
\label{sec:atmosphere}
Atmospheric fluctuations generate scale-dependent noise. A frozen pattern of 2D fluctuations blowing across the field of view at fixed height above the instrument produces a $1/f$ or Kolmogorov spectrum with an approximate form of $P_{\mathrm{N}}(\ell) \propto \ell^{-8/3}$. Here we have introduced $\ell$, the angular counterpart to $k$, which is the Fourier transform pair of the angle $\theta$ on the sky. Since atmospheric fluctuations are local to the instrument, it is common to express their effects in $\ell$ rather than through the redshift dependent mapping to k. Since the fluctuations are finite in size, they only affect the largest accessible scales, with the cut-off used to define the parameter $\ell_{\mathrm{knee}}$, such that 
\begin{equation}
\label{eqn:lknee}
P_{\mathrm{N}}(\ell) = P_{\mathrm{N}} \left(1 + \left(\frac{\ell}{\ell_{\rm knee}} \right)^{\alpha}\right). 
\end{equation}

The values of $\ell_{\rm knee}$ and $\alpha$ are determined empirically from fits to observed band powers at fixed scan rate. In our forecasts we fix $\alpha = -2.8$ and $\ell_{\rm knee} = 200$ to be consistent with measured values from contemporary fast-scanning CMB experiments in temperature \citep{Ade_2018}. These values are scan strategy-dependent and should be viewed as approximate. Moreover, it is possible that LIM measurements will have improved noise properties due to the ability to excise atmospheric lines in the spectroscopic measurement. Near-future pathfinder experiments will provide more detailed atmospheric characterization suitable for LIM forecasts and better inform estimates of the largest scales at which LIM is sensitive to cosmology.

\subsubsection{Interloper Lines}
\label{sec:interlopers} 

For observations at fixed redshift and target line frequency, line confusion arises because the emission from sources at various redshifts overlaps in observed frequency. Without additional information, observed power at a given target redshift cannot be easily distinguished from power at a different redshift that has the same observed frequency. This effect can be large. For example, for [CII] experiments targeting $z \sim 7$, CO rotational transitions between $z=0.45$ and $z=1.8$ act as foregrounds with power larger than the target line. For the low-$J$ transitions of CO that we target, higher rotational transitions are the main source of interloper confusion. To model this scale-dependent effect, we modify the numerator of Eq.~\ref{Eqn: variance} to
\begin{equation}
    \label{eqn:interloper} 
    \Sigma_{i} b_i(z) I_i(z) P_{\mathrm{m}}(k,z_i) + P_{\mathrm{shot}} + P_{\mathrm{N}}.
\end{equation}

For the RSD multipoles, we similarly modify Eq.~\ref{eqn:multi_cov} to sum over the RSD power spectrum at each redshift, 
\begin{multline}
\label{eqn:multi_cov_interloper}
    \mathrm{Cov}_{l, l'}(k) = \frac{(2l + 1)(2l' + 1)}{N_{m}} \\ \times\int_{-1}^{1} \mathscr{L}_l(\mu)\mathscr{L}_{l'}(\mu) (\Sigma_i P_{\mathrm{m}}(k, \mu, z_i) + P_{\mathrm{N}})^2 \ d\mu. 
\end{multline}

In other words, we assume that the interloper power adds to the noise, and does not contribute signal to the estimate of $P_{\mathrm{m}}(k, z_i)$. In fact, the interloper contributions themselves carry cosmological information similar to the information that we will consider from the brightest lines. For example, in a wideband experiment observing a large range in redshift and different rotational CO transitions simultaneously, internal cross-correlations may be able to extract the underlying matter power spectrum from each CO line and add them coherently to the signal. 

A wide variety of techniques has been proposed for interloper deconfusion. For surveys targeting higher redshifts, masking techniques---i.e., removing brighter pixels that are more likely to come from lower redshifts \cite{Yue2015, 2015MNRAS.452.3408B}, or using an external catalog of bright interloping galaxies \cite{sun2018}) can significantly reduce the interloper contribution. Cross-correlations between lines can also reconstruct a high percentage of the true underlying map \citep{Chung2019, Cheng_2020}. Finally, geometric tests for interloper deconfusion were introduced in \cite{silva2015prospects} and \cite{Lidz_2016}.

\subsubsection{Galactic Continuum Foregrounds}

Galactic continuum emission can be a significant foreground for both CMB and LIM experiments. For CO, thermal dust can significantly eclipse the line brightness temperatures at frequencies above 50 GHz while non-thermal synchrotron emission dominates at lower frequencies. By fitting a smooth, low-order polynomial to a foreground that slowly varies in frequency, this can be subtracted and removed, leaving only the underlying cosmological signal. However, residuals from fitting these broadband terms can lead to spuriously inferred excess matter power at large scales that is a function of the residuals after continuum subtraction \citep{McQuinn_2006}.

We used NBODYKIT to combine the linear matter power spectrum with the Galactic dust continuum, and studied recovery of the power spectrum. We began by adding a mock LIM signal to a typical Galactic dust spectrum, and then removed a series of low-order polynomials. We find that even under pessimistic assumptions, foreground removal only affects large scales ($k \sim 10^{-3}$ h/Mpc)  that contribute little weight to the overall constraint (due to the small number of available modes), and which are additionally impacted by atmospheric noise. In a simple but more realistic model for the continuum fitting, the difference between the input and recovered spectrum is less than 1$\%$ on intermediate scales. We therefore neglect galactic continuum foregrounds in our forecast.

\section{Results} 
\label{sec:results}

In this section, we review the Fisher Matrix formalism used to derive constraints, and describe the specifics of the survey we forecast, motivated by the relevant scales needed to constrain modified gravity. We then present constraints for a future mm-wave LIM experiment as a function of sensitivity, and account for various systematic effects.

\subsection{Fisher Matrix Formalism} 
\label{sec:fisher}

Fisher matrix methods are a standard way of estimating the precision of future experiments \citep{2006astro.ph..9591A}. Beginning from an assumption of Gaussian errors, by Taylor expanding about the true parameter values, we have 
\begin{equation}
    \label{Fisher1} 
    \exp
    \left(-\frac{1}{2} \chi^2\right) \propto \exp\left(-\frac{1}{2}F_{jk} \delta p_j \delta p_k\right)
\end{equation}
where the matrix $F_{jk}$ is called the Fisher matrix, and can be evaluated as
\begin{equation}
    \label{Fisher2} 
    F_{jk} = \sum_b \frac{N_b}{\sigma_b^2}\frac{\delta f_b}{\delta p_j}\frac{\delta f_b}{\delta p_k}.
\end{equation}

The Fisher matrix is equivalent to the inverse of the covariance matrix. Equation~\ref{Fisher2} instructs us to estimate the covariance matrix by computing derivatives of the observable quantity in bins labelled by $b$ and with corresponding error $\sigma_b$. Inverting $F_{jk}$ then yields the variance and covariance of the model parameters. In our case, we use the binned power spectrum, $P_b(k)$, and estimate the error per bin $\sigma_b$ using Equations~\ref{Eqn: variance}, \ref{smoothing}, and \ref{eqn:interloper}. An explicit expression of the Fisher matrix in terms of the modified gravity parameters is
\begin{align}
F_{M,B} = \sum_k \frac{N_k}{\sigma_k^2}\frac{\delta P(k)}{\delta c_{M,B}}\frac{\delta P(k)}{\delta c_{M,B}} .
\end{align}

Equation~\ref{Fisher2} is sufficient for estimating the covariance in the $c_B$ and $c_M$ parameters from a single measurement of the power spectrum. In the case where multiple emission lines are independently used to constrain the power spectrum shape, the combined Fisher matrix is given by the sum of the independent Fisher matrices for each line, $F^{A+B} = F^A + F^B$. 

As the RSD multipole moments are not statistically independent, we compute the joint constraint from the full covariance matrix: 

\begin{align}
    \label{explicit_fisher}
F_{M,B} = \sum_k \sum_{l, l'} \frac{\delta P_l(k)}{\delta c_{M,B}} C_{l, l'}\frac{\delta P_{l'}(k)}{\delta c_{M,B}}, 
\end{align}

where l, l' run over the 0 and 2 RSD multipole moments. 

\subsection{Survey Definition and Accessible Scales} 
\label{sec:survey}

Constraints on the modified gravity models considered here will search for a \textit{nearly scale-invariant change} in the matter power spectrum from $\Lambda$CDM for $k \gtrsim 10^{-2}$ h/Mpc.  This implies that the astrophysical line emission terms in Eq.~\ref{eqn:Pclust} need to be known to better than the $\sim$few \% deviations in $P_{\mathrm{m}}$ that we are considering. A LIM survey's sensitivity to the power spectrum falls off at the largest scales due to foreground filtering, atmospheric noise, and the decreasing number of Fourier modes in a finite survey volume. A heuristic for the sensitivity of a survey to an observable, e.g. the matter power spectrum on a given $k$-scale, is to count the number of accessible modes accessible at that scale.

The number of observable modes can be improved by increasing either the spectral or angular resolution or survey sky fraction. The LIM surveys we consider here are mismatched in angular and spectral resolution; while the arcminute scales accessible with 5--10m class dishes correspond to $k\sim 1-10$ h/Mpc, current mm-wave spectrometers  have only been demonstrated up to $R\sim 300$ corresponding to $k\sim $ 0.2 . However, a factor of $\sim$several improvement in resolution should be possible with technology developments in the near future. In Fig.~\ref{fig: relevant_scales} we show the matter power spectrum ($k^3 P_{\mathrm{m}}(k)$) and cumulative signal-to-noise ratio on the power spectrum deviation as a function of $k$ for representative values of $c_M, c_B$ in both parameterizations at $z=0.5$, focusing on the difference between the $R = \frac{\delta\nu}{\nu} = 300$ and $1000$ cases. Increasing the spectral resolution increases the  number of modes in the survey, leading to improved sensitivity even when a survey is unable to resolve the smallest scale structures. Most of the constraining power in the $R=300$ case occurs around $k \sim 0.1$ h/Mpc, due to the larger number of modes after accounting for both the spectral resolution and number of modes contained in the survey volume. 
The scale at which non-linear growth affects the power spectrum at the $2\%$ level is $k \sim 0.1$ h/Mpc at $z \approx 0$, and $k \sim 0.25$ h/Mpc at $z=3$ (the mean redshift of the lines considered is shown in Table~\ref{tab:targets}). Thus for $R=1000$ about $50\%$ of the constraining power comes from weakly non-linear scales. The exact level of non-linear growth expected in Horndeski gravity is uncertain, but experiments with \texttt{HALOFIT} \citep{2003MNRAS.341.1311S} suggest a moderate increase in the matter power spectrum, and thus moderately improved sensitivity. To be conservative, we assume the predictions of linear theory. Another source of non-linear biasing is the relationship between CO emissivity and dark matter power, which depends on the CO luminosity function \citep{Breysse2014} and exhibits non-linear effects at $k \sim$ $0.2$ h/Mpc \citep{Einasto2019}. The remaining uncertainties in the scale at which nonlinear biasing becomes important will be decreased with future small-scale detection of CO shot noise.

As a baseline survey definition, we consider a survey over 40$\%$ of the sky observing 75--310 GHz with $R = 300$. The sky fraction is set in part by the physical limits of telescopes and optics that often restrict observing to elevation angles $\ge$ 40-50 deg. Bright emission from the Galactic center can further restrict accessible sky fractions by another $\approx$ 10$\%$. This survey geometry corresponds to a range of accessible scales between $\approx$ $2 \times 10^{-3} \le k \le 5 \times 10^{-1}$ h/Mpc. The maximum scale is set by the resolution in the frequency direction while the minimum scale is set by the assumed sky fraction. Increasing the sky fraction to $70\%$ improves access to the largest scales by about a factor of two, while the smallest scales remain limited by the resolution in the frequency direction. Atmospheric and galactic thermal continuum foregrounds can also limit sensitivity to the largest scale modes.

Fixing the sky fraction and bandwidth allows us to make the estimates of the noise power given in Table~\ref{tab:targets}. The white noise contribution arises from incident photon power from the atmosphere, telescope, and detector (Equations \ref{eqn: NEP}, \ref{eqn: Q}). Within each of the mm-wave atmospheric windows ($\approx$ 75--115, 125--175, and 180--310 GHz), we use an effective NET in which all of the frequency channels within each band are nverse variance weighted. We then calculate the voxel volume using Equation \ref{eqn: V_vox} for the minimum spatial and frequency scales, set by the telescope's angular resolution and spectrometer spectral resolution, respectively. We convert between NET and integration time via Equation \ref{eqn: Pn}.

\subsection{Fiducial Analysis}  

\begin{figure*}
\includegraphics[scale=0.33]{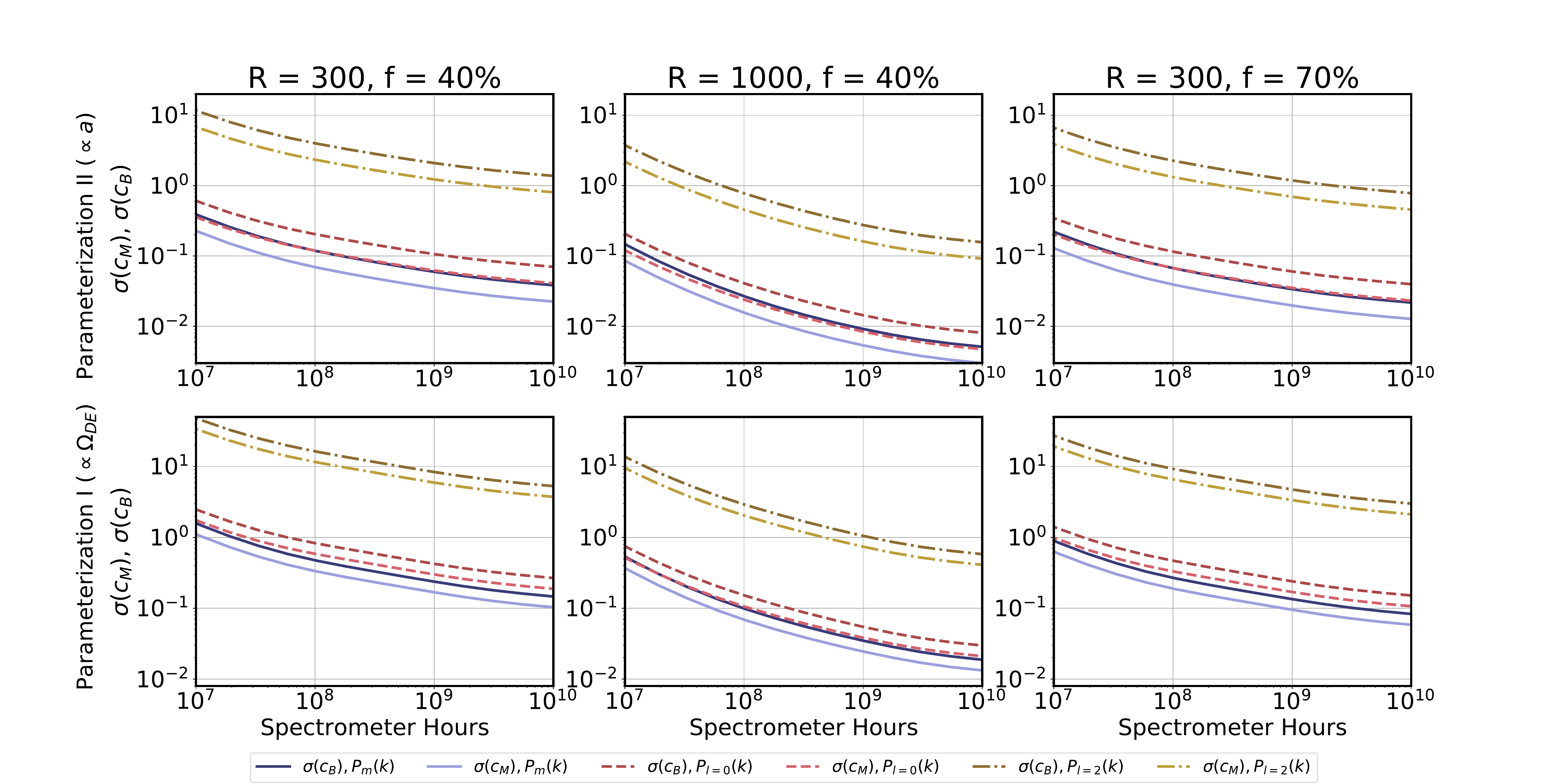}
\caption{\label{fig:baseline} Sensitivity to the $c_M, c_B$ parameters from the matter power spectrum or redshift space distortion monopole differs by a factor of $\approx 2$ independent of spectral resolution or sky fraction.  Here we show forecasted sensitivity (posterior width) as a function of spectrometer hours for the $c_M$ and $c_B$ parameters in the baseline ($R=300$, $f_{\mathrm{sky}} = 40\%$), increased spectral resolution ($R=1000$, $f_{\mathrm{sky}} = 40\%$), and increased survey volume ($R=300$, $f_{\mathrm{sky}} = 70\%$) cases. Top panels are for Parameterization II and the bottom panel is for Parameterization I.}
\end{figure*}

For our fiducial survey, we assume the experiment described in the previous section, with $R=300$ spectral resolution and $f_{\mathrm{sky}} = 40\%$. In Table \ref{tab:targets} we summarize our target lines, redshifts, and shot noise estimates. Combined constraints are obtained by summing the Fisher matrices over the full bandwidth, assuming statistical independence of each target line and redshift. We focus on the effects of survey definition for the sensitivity of the baseline survey to $c_M, c_B$, leaving the effects of interlopers and atmospheric foregrounds for the next section.

Figure \ref{fig:baseline} shows the 1$\sigma$ posterior widths, $\sigma(c_b)$ and $\sigma(c_M)$, in Parameterization II  ($\alpha_i \propto a$, top row) and Parameterization I ($\alpha_i \propto \Omega_{\Lambda}$, bottom row), as a function of \textit{spectrometer-hours}, the product of the number of spectrometers and integration time. In addition to the matter power spectrum constraint, we also show the results from the RSD power spectrum monopole and quadrupole separately. The fiducial experiment ($f_{\mathrm{sky}} = 40\%, R = 300)$ achieves similar sensitivity to the RSD monopole and matter power spectrum, while the quadrupole is significantly less constraining \citep{Chung2019}. 

Sensitivity can be improved by increasing the number of modes through larger survey volumes or improved spectral resolution. In the center and right panels of Figure \ref{fig:baseline}, we forecast for increasing the spectral resolution from $R=300$ to $R=1000$ at fixed sky fraction, and for increasing the sky fraction from $f_{\mathrm{sky}} = 40\%$ to $f_{\mathrm{sky}} = 70\%$. We find a factor of a few improvement in sensitivity, achieving a $\pm 0.1$ level constraint on each of the $c_B, c_M$ in Parameterization II at $10^8$ spectrometer-hours, with improved sensitivity with longer integration times in both the $R = 1000$ and $f_{\mathrm{sky}} = 70\%$ experiments.

Increasing the spectral resolution provides larger returns on sensitivity than going to higher sky fractions at fixed spectrometer-hours, with sensitivity approaching the $\pm 0.01$ level in $\approx 10^8$ spectrometer hours. However, this result depends on the assumed shot noise for each target line. In a test where the shot noise was assumed to take its $z=2$ values from \citet{dizgah2021neutrino}, sensitivity saturates near $\pm 0.1$ at $10^8$ spectrometer-hours in the $R=1000$ experiment in both parameterizations. Less sensitivity to the $c_M$, $c_B$ is achieved in Parameterization I, regardless of survey definition.

\begin{figure*}
\includegraphics[scale=0.12]{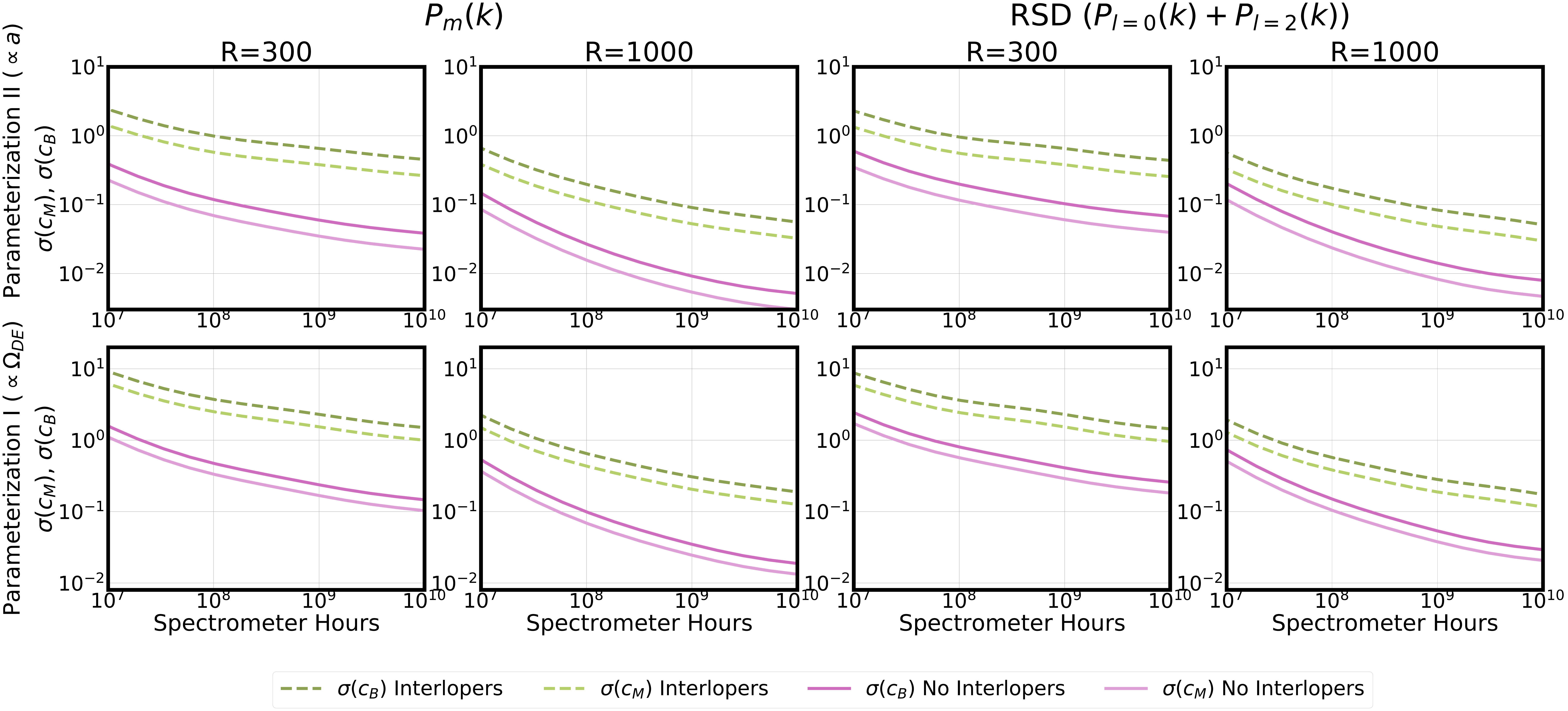}
\caption{\label{fig:interlopers} Including interlopers leads to a decrease in sensitivity both from $P_{\mathrm{m}}(k)$ and the sum of $P_{l=0}(k) + P_{l=2}(k)$, with a reduced sensitivity gap for the redshift space measurements. As before, we plot forecasted sensitivity (posterior width) as a function of spectrometer hours for the $c_M$ and $c_B$ parameters when interlopers are included or excluded in the baseline survey. Left panels show sensitivity from the matter power spectrum, right panels for RSD multipoles. Top panels show Parameterization II, bottom panels show Parameterization I.}
\end{figure*}

\subsection{Accounting for Interlopers and Low-Frequency Noise}

In order to quantify the effect of different analysis choices on sensitivity to the power spectrum, we now consider the impact that interlopers and low-frequency noise have on measurements of the $c_M$ and, $c_B$ parameters. We begin by modifying the fiducial analysis and baseline survey according to the discussion in Section \ref{sec:noise}. Both the survey geometry and atmospheric scale limit the maximum accessible scales. Since the signal to noise ratio on measurements of the power spectrum decreases significantly on the largest scales, the atmospheric parameters $\ell_\knee$ and $\alpha$ will only significantly impact the constraint if they differ substantially from the scale set by the survey geometry. As discussed in Section~\ref{sec:atmosphere}, we assume that the atmospheric noise will be similar to that observed at the South Pole. As the relevant observable scales are above $\ell_{\knee}$ (see Figure \ref{fig: relevant_scales}) the choice of $\ell_{\knee}$ has little impact on our results.

We further consider the effect of interloper lines that mimic redshift dependent intensity fluctuations and therefore pose a potentially more serious problem. Interlopers can mimic a modified gravity effect, since at fixed redshift a change in the intensity bias is degenerate with a change in the growth function. In Figure~\ref{fig:interlopers}, we plot the sensitivity as a function of spectrometer-hours for both the baseline case ($f_{\mathrm{sky}} = 40\%, R = 300$) and a case including interlopers and low frequency noise. We treat the interlopers following Section \ref{sec:interlopers}, where interloper lines are assumed to contribute noise but not signal to the measurement of the modified gravity parameters. Here we consider the sensitivity from $P_{\mathrm{m}}(k)$ and the sum of $P_{l = 0}(k)$ and $P_{l = 2}(k)$ computed using Equations \ref{Cl00} to \ref{Cl22} and \ref{explicit_fisher}. Interlopers are treated as in Equation \ref{eqn:multi_cov_interloper}.

The inclusion of interlopers can significantly reduce constraining power. While we obtain $\pm 0.1$ level constraints in the fidicial survey for Parameterization II in $10^8$ spectrometer hours, this now requires $10^9$ spectrometer hours or more. In Parameterization I, a $\pm 0.1$ constraint is no longer obtained in our range of spectrometer-hours. While such a measurement would still allow for characterization of the size of modified gravity effect over a range in redshift, an interloper-contaminated LIM measurement would add only a very limited amount of information as compared to the existing CMB and LSS measurements.

When there is clean separation between interloper and target lines (the ``interloper free'' baseline case), $P_{\mathrm{m}}(k)$ is more sensitive to the values of the $c_M, c_B$ than the sum of the information from $P_{l=0}(k)$ and $P_{l=2}(k)$. However, in the case of poor line separation, this situation is reversed, where the RSD multipole moments retain more of the sensitivity that is lost in the matter power spectrum. That is, the difference between the interloper and interloper-free cases is smaller. This result is anticipated by the close relationship between Alcock-Paczynski tests, which can be used to achieve line separation, and the redshift space distortion. The two effects are degenerate with the matter power spectrum, and require an assumed background cosmology to fully isolate from one another \citep{Samushia_2012}. The degree of line seperation working in redshift space depends on the linear growth factor $f$ and ratios of volumes between the target and interloper line redshifts.

The two cases we have considered here (foreground/interloper-free and interloper-contaminated) roughly bound the range of expected sensitivity. The interloper-contaminated case we have considered is unrealistically pessimistic, where no attempt is made to remove interlopers before performing a cosmological analysis. Numerous techniques have been proposed in the literature for reducing their contributions (see Section~\ref{sec:interlopers}). Although outside the scope of this work, one complication for LIM measurements of modified gravity is that several interloper mitigation schemes (e.g., geometric methods) depend on an assumed near-$\Lambda$CDM expansion history. Fully quantifying the effect of various assumptions on the recovery of the signatures of modified gravity is left for future work. 

\section{Discussion}
\label{sec:discussion}

The observable signature of Horndeski gravity on LSS is a scale-independent change in the normalisation of the matter power spectrum for $k > 10^{-3}$ h/Mpc, observable in either comoving or redshift space. 

\cite{Noller_2019} found that the inclusion of $f\sigma_8$ from BOSS DR11 CMASS and 6dF led to increased sensitivity relative to the CMB-only and CMB + mPk cases. We therefore also forecast for an experiment targeting the RSD monopole and quadrupole, which carry information about the velocity field. We find similar sensitivity to $c_B, c_M$ from combining the first two moments of the RSD power spectrum and the matter power spectrum alone. Consistent with the expectations from \citet{Chung2019}, the quadrupole contributes limited sensitivity compared to the monopole-only result. This is because the uncertainties on the quadrupole power spectrum are a factor of $\approx$ $\sqrt{2l + 1}/2$ larger than the uncertainties on the monopole. We assume fiducial models for the line biases and temperatures as discussed in Section \ref{sec:models} under the assumption that both will be well constrained by future experiments, for example, through multiple-line cross correlations or via cross-correlation with galaxy surveys \cite{Chung2019}.

Unmitigated interloper emission can reduce the sensitivity at fixed integration time by roughly an order of magnitude. This is expected since the primary effect of modifying gravity in our parameterizations mimics a change to the growth function with redshift. As interloper lines add noise power from a range of redshifts, this adds scatter to the inferred growth function, or equivalently the overall amplitude of the power spectrum on a range of scales.

The inclusion of interlopers in our baseline surveys leads to a reduction in sensitivity to the Horndeski linear theory $\alpha$ functions. We expect the sensitivity of a future experiment to lie somewhere between the no interlopers and interlopers cases shown in Figure \ref{fig:interlopers}. Although a large number of methods to mitigate the effect of interlopers on cosmological analyses have been studied in the literature, this motivates future work to understand how interlopers may bias future measurements in cosmology. 

A LIM experiment will produce a set of redshift-dependent power spectrum amplitudes that are weighted by the line temperatures and bias factors. Under a fixed background cosmology and assumed evolution of the line intensities, internal and external cross correlations can be used to both disentangle the interloper contributions and significantly reduce degeneracies between the astrophysics- and cosmology-dependent terms. 

Our constraints make use of only the average line intensity across the target band and not its redshift evolution. However, the line intensity is expected to trace star formation and therefore peak at $z\approx 2$, while the modified gravity power spectrum excess is expected to grow monotonically with redshift. Therefore, the evolution of the two effects is expected to generically differ with an overall change in the line evolution as compared to $\Lambda$CDM. This provides another potential avenue for a LIM experiment to probe modified gravity directly from the redshift evolution.  Making use of this information will require improvements in our understanding of the line evolution ($I(z)$) models and scaling relations that link these models to the SFR.

While direct constraints on Horndeski gravity from galaxy surveys have been challenged by limited cosmological volumes and uncertainty in the galaxy bias, next-generation galaxy surveys will probe larger volumes, allowing for joint analysis and cross-correlations that can break degeneracies between multiple probes. Galaxy-LIM cross correlations and multi-line LIM cross correlations, for example, can separate the line bias and intensities even in the presence of interlopers \citep{Schaan_2021}.

\section{Conclusions}
\label{sec:conclusion}

In this work, we investigated the ability of a wide-bandwidth ground-based LIM experiment targeting rotational CO transitions to constrain the linear theory parameters of Horndeski models. We consider two parameterizations for the evolution of these parameters, governing the braiding and running of the Planck mass, where both are allowed to evolve with the effective dark energy density $\Omega_{DE}$ or with the scale factor $a$. Both parameterizations predict larger effects at low redshift, with excesses in apparent power at small scales and deficits in power at large scales for a large part of this 2D parameter space. 

With observations in three atmospheric bands from 75--310 GHz, we find that the bright rotational CO transitions from redshifts 0--3 yield posterior widths for these parameters approaching the sensitivity of CMB and existing galaxy survey constraints at 10$^8$--10$^9$ spectrometer-hours. This result is robust to the presence of continuum foreground and atmospheric effects, being primarily driven by information obtained from intermediate scales and therefore mainly limited by the degree of interloper contamination. Models in which the modified gravity effect is proportional to the scale factor rather than $\Omega_{DE}$ yield constraints that are about an order of magnitude larger at fixed integration time, a result that is consistent with past measurements. There is significant uncertainty about what limits the sensitivity of future experiments in the space of noise, astrophysical, and cosmological modeling uncertainties, and our results should therefore be viewed as a preliminary estimate of the performance of a real instrument. Nonetheless, these results show that future LIM experiments could place competitive constraints on the space of modified gravity theories. 

Horndeski theories represent a general class of modified gravity models that add scalar-coupled terms to the gravitational Lagrangian. As discussed in \cite{Bellini_2014}, measuring values of the $\alpha$ functions therefore constrains the parameter space of viable modifications to General Relativity.  These include metric f(R), Kinetic Gravity Braiding, Galileon, Brans-Dicke, Palantini, and Gauss-Bonnet models.

LIM experiments with CMB heritage could potentially reach 10$^8$--10$^9$ spectrometer hours over the next 10--15 years. On similar timescales, space-based spectro-polarimeters operating in the far-IR are expected to become feasible \citep{delabrouille2019}. A space-based instrument would trade angular spatial resolution for increased sensitivity to the integrated line emission through wider bandwidth and reduced large scale noise due to a lack of atmosphere. Combined with a larger $f_{\mathrm{sky}}$, this would enable a range of complementary CMB and galaxy cluster science. As the signature of a modified gravity effect is largely scale-independent on intermediate scales, such an experiment would be able to improve constraints on deviations from General Relativity through both direct measurement of the matter power spectrum and through multi-tracer analyses similar to the one we consider here.

Measurement of modified gravity effects will require improvements in our knowledge of target line biases and intensities to break parameter degeneracies. While analysis and modeling methods for LIM remain in their infancy compared to well developed-methods for CMB and galaxy survey measurements, LIM experiments targeting rotational CO benefit from both this heritage and bright line temperatures. This makes these transitions promising targets for constraining modified gravity theories. Our results show that future LIM experiments can achieve constraints on the linear parameters of Horndeski theories that are competitive with the current state of the art. 

\section*{Acknowledgements}

We thank Garrett Keating for shot noise estimates used in this forecast and for helpful conversations. We also thank Johannes Noller for helpful comments on both constraints from gravitational waves and the BOSS power spectrum normalization.

SB is supported by NSF grant AST-1817256. KSK is supported by an NSF Astronomy and Astrophysics Postdoctoral Fellowship under award AST-2001802.

%%%%%%%%%%%%%%%%%%%%%%%%%%%%%%%%%%%%%%%%%%%%%%%%%%
\section*{Data Availability}

All data is available publicly at 
https://github.com/bscot/Horndeski-LIM



\bibliographystyle{mnras}
\bibliography{example} % if your bibtex file is called example.bib

\begin{thebibliography}{}
\makeatletter
\relax
\def\mn@urlcharsother{\let\do\@makeother \do\$\do\&\do\#\do\^\do\_\do\%\do\~}
\def\mn@doi{\begingroup\mn@urlcharsother \@ifnextchar [ {\mn@doi@}
  {\mn@doi@[]}}
\def\mn@doi@[#1]#2{\def\@tempa{#1}\ifx\@tempa\@empty \href
  {http://dx.doi.org/#2} {doi:#2}\else \href {http://dx.doi.org/#2} {#1}\fi
  \endgroup}
\def\mn@eprint#1#2{\mn@eprint@#1:#2::\@nil}
\def\mn@eprint@arXiv#1{\href {http://arxiv.org/abs/#1} {{\tt arXiv:#1}}}
\def\mn@eprint@dblp#1{\href {http://dblp.uni-trier.de/rec/bibtex/#1.xml}
  {dblp:#1}}
\def\mn@eprint@#1:#2:#3:#4\@nil{\def\@tempa {#1}\def\@tempb {#2}\def\@tempc
  {#3}\ifx \@tempc \@empty \let \@tempc \@tempb \let \@tempb \@tempa \fi \ifx
  \@tempb \@empty \def\@tempb {arXiv}\fi \@ifundefined
  {mn@eprint@\@tempb}{\@tempb:\@tempc}{\expandafter \expandafter \csname
  mn@eprint@\@tempb\endcsname \expandafter{\@tempc}}}

\bibitem[\protect\citeauthoryear{{Abazajian} et~al.,}{{Abazajian}
  et~al.}{2016}]{Abazajian2016}
{Abazajian} K.~N.,  et~al., 2016, arXiv e-prints, \href
  {https://ui.adsabs.harvard.edu/abs/2016arXiv161002743A} {p. arXiv:1610.02743}

\bibitem[\protect\citeauthoryear{Ade et~al.,}{Ade et~al.}{2018}]{Ade_2018}
Ade P.,  et~al., 2018, \mn@doi [Physical Review Letters]
  {10.1103/physrevlett.121.221301}, 121

\bibitem[\protect\citeauthoryear{Ade et~al.,}{Ade et~al.}{2020}]{Ade2020}
Ade P.,  et~al., 2020, \mn@doi [Astronomy & Astrophysics]
  {10.1051/0004-6361/202038456}, 642, A60

\bibitem[\protect\citeauthoryear{Aghanim et~al.,}{Aghanim
  et~al.}{2020}]{Planck_2020}
Aghanim N.,  et~al., 2020, \mn@doi [Astronomy & Astrophysics]
  {10.1051/0004-6361/201833910}, 641, A6

\bibitem[\protect\citeauthoryear{{Albrecht} et~al.,}{{Albrecht}
  et~al.}{2006}]{2006astro.ph..9591A}
{Albrecht} A.,  et~al., 2006, arXiv e-prints, \href
  {https://ui.adsabs.harvard.edu/abs/2006astro.ph..9591A} {pp
  astro--ph/0609591}

\bibitem[\protect\citeauthoryear{{Arai} \& {Nishizawa}}{{Arai} \&
  {Nishizawa}}{2018}]{2018PhRvD..97j4038A}
{Arai} S.,  {Nishizawa} A.,  2018, \mn@doi [\prd] {10.1103/PhysRevD.97.104038},
  \href {https://ui.adsabs.harvard.edu/abs/2018PhRvD..97j4038A} {97, 104038}

\bibitem[\protect\citeauthoryear{{BICEP2 Collaboration} et~al.,}{{BICEP2
  Collaboration} et~al.}{2015}]{bicep_dets}
{BICEP2 Collaboration} et~al., 2015, \mn@doi [\apj]
  {10.1088/0004-637X/812/2/176}, \href
  {https://ui.adsabs.harvard.edu/abs/2015ApJ...812..176B} {812, 176}

\bibitem[\protect\citeauthoryear{{Baker}, {Bellini}, {Ferreira}, {Lagos},
  {Noller}  \& {Sawicki}}{{Baker} et~al.}{2017}]{2017PhRvL.119y1301B}
{Baker} T.,  {Bellini} E.,  {Ferreira} P.~G.,  {Lagos} M.,  {Noller} J.,
  {Sawicki} I.,  2017, \mn@doi [\prl] {10.1103/PhysRevLett.119.251301}, \href
  {https://ui.adsabs.harvard.edu/abs/2017PhRvL.119y1301B} {119, 251301}

\bibitem[\protect\citeauthoryear{Bellini \& Sawicki}{Bellini \&
  Sawicki}{2014}]{Bellini_2014}
Bellini E.,  Sawicki I.,  2014, \mn@doi [Journal of Cosmology and Astroparticle
  Physics] {10.1088/1475-7516/2014/07/050}, 2014, 050–050

\bibitem[\protect\citeauthoryear{Bellini, Sawicki  \& Zumalacárregui}{Bellini
  et~al.}{2020}]{Bellini_2020}
Bellini E.,  Sawicki I.,   Zumalacárregui M.,  2020, \mn@doi [Journal of
  Cosmology and Astroparticle Physics] {10.1088/1475-7516/2020/02/008}, 2020,
  008–008

\bibitem[\protect\citeauthoryear{Bernal \& Kovetz}{Bernal \&
  Kovetz}{2022}]{Bernal_2022}
Bernal J.~L.,  Kovetz E.~D.,  2022, Line-Intensity Mapping: Theory Review,
  \mn@doi{10.48550/ARXIV.2206.15377}, \url {https://arxiv.org/abs/2206.15377}

\bibitem[\protect\citeauthoryear{Blas, Lesgourgues  \& Tram}{Blas
  et~al.}{2011}]{Blas_2011}
Blas D.,  Lesgourgues J.,   Tram T.,  2011, \mn@doi [Journal of Cosmology and
  Astroparticle Physics] {10.1088/1475-7516/2011/07/034}, 2011, 034–034

\bibitem[\protect\citeauthoryear{Breysse, Kovetz  \& Kamionkowski}{Breysse
  et~al.}{2014}]{Breysse2014}
Breysse P.~C.,  Kovetz E.~D.,   Kamionkowski M.,  2014, \mn@doi [Monthly
  Notices of the Royal Astronomical Society] {10.1093/mnras/stu1312}, 443,
  3506–3512

\bibitem[\protect\citeauthoryear{{Breysse}, {Kovetz}  \&
  {Kamionkowski}}{{Breysse} et~al.}{2015}]{2015MNRAS.452.3408B}
{Breysse} P.~C.,  {Kovetz} E.~D.,   {Kamionkowski} M.,  2015, \mn@doi [\mnras]
  {10.1093/mnras/stv1476}, \href
  {https://ui.adsabs.harvard.edu/abs/2015MNRAS.452.3408B} {452, 3408}

\bibitem[\protect\citeauthoryear{Breysse, Yang, Somerville, Pullen, Popping  \&
  Maniyar}{Breysse et~al.}{2021}]{breysse2021estimating}
Breysse P.~C.,  Yang S.,  Somerville R.~S.,  Pullen A.~R.,  Popping G.,
  Maniyar A.~S.,  2021, On estimating the cosmic molecular gas density from CO
  Line Intensity Mapping observations (\mn@eprint {arXiv} {2106.14904})

\bibitem[\protect\citeauthoryear{Cheng, Chang  \& Bock}{Cheng
  et~al.}{2020}]{Cheng_2020}
Cheng Y.-T.,  Chang T.-C.,   Bock J.~J.,  2020, \mn@doi [The Astrophysical
  Journal] {10.3847/1538-4357/abb023}, 901, 142

\bibitem[\protect\citeauthoryear{Chung et~al.,}{Chung et~al.}{2019}]{Chung2019}
Chung D.~T.,  et~al., 2019, \mn@doi [The Astrophysical Journal]
  {10.3847/1538-4357/ab0027}, 872, 186

\bibitem[\protect\citeauthoryear{Clifton, Ferreira, Padilla  \&
  Skordis}{Clifton et~al.}{2012}]{Clifton_2012}
Clifton T.,  Ferreira P.~G.,  Padilla A.,   Skordis C.,  2012, \mn@doi [Physics
  Reports] {10.1016/j.physrep.2012.01.001}, 513, 1–189

\bibitem[\protect\citeauthoryear{{Cosmic Visions 21 cm Collaboration}
  et~al.,}{{Cosmic Visions 21 cm Collaboration}
  et~al.}{2019}]{cosmicvisions21cmcollaboration2019inflation}
{Cosmic Visions 21 cm Collaboration} et~al., 2019, Inflation and Early Dark
  Energy with a Stage II Hydrogen Intensity Mapping Experiment (\mn@eprint
  {arXiv} {1810.09572})

\bibitem[\protect\citeauthoryear{{Creminelli}, {Tambalo}, {Vernizzi}  \&
  {Yingcharoenrat}}{{Creminelli} et~al.}{2020}]{2020JCAP...05..002C}
{Creminelli} P.,  {Tambalo} G.,  {Vernizzi} F.,   {Yingcharoenrat} V.,  2020,
  \mn@doi [\jcap] {10.1088/1475-7516/2020/05/002}, \href
  {https://ui.adsabs.harvard.edu/abs/2020JCAP...05..002C} {2020, 002}

\bibitem[\protect\citeauthoryear{Creque-Sarbinowski \&
  Kamionkowski}{Creque-Sarbinowski \& Kamionkowski}{2018}]{Creque2018}
Creque-Sarbinowski C.,  Kamionkowski M.,  2018, \mn@doi [Physical Review D]
  {10.1103/physrevd.98.063524}, 98

\bibitem[\protect\citeauthoryear{{Crites} et~al.,}{{Crites}
  et~al.}{2014}]{2014SPIE.9153E..1WC}
{Crites} A.~T.,  et~al., 2014, in {Holland} W.~S.,  {Zmuidzinas} J.,  eds,
  Society of Photo-Optical Instrumentation Engineers (SPIE) Conference Series
  Vol. 9153, Millimeter, Submillimeter, and Far-Infrared Detectors and
  Instrumentation for Astronomy VII. p. 91531W, \mn@doi{10.1117/12.2057207}

\bibitem[\protect\citeauthoryear{DeBoer et~al.,}{DeBoer
  et~al.}{2017}]{DeBoer_2017}
DeBoer D.~R.,  et~al., 2017, \mn@doi [Publications of the Astronomical Society
  of the Pacific] {10.1088/1538-3873/129/974/045001}, 129, 045001

\bibitem[\protect\citeauthoryear{{Delabrouille} et~al.,}{{Delabrouille}
  et~al.}{2019}]{delabrouille2019}
{Delabrouille} J.,  et~al., 2019, arXiv e-prints, \href
  {https://ui.adsabs.harvard.edu/abs/2019arXiv190901591D} {p. arXiv:1909.01591}

\bibitem[\protect\citeauthoryear{Dizgah, Keating, Karkare, Crites  \&
  Choudhury}{Dizgah et~al.}{2021}]{dizgah2021neutrino}
Dizgah A.~M.,  Keating G.~K.,  Karkare K.~S.,  Crites A.,   Choudhury S.~R.,
  2021, Neutrino Properties with Ground-Based Millimeter-Wavelength Line
  Intensity Mapping (\mn@eprint {arXiv} {2110.00014})

\bibitem[\protect\citeauthoryear{{Einasto}, {Liivam{\"a}gi}, {Suhhonenko}  \&
  {Einasto}}{{Einasto} et~al.}{2019}]{Einasto2019}
{Einasto} J.,  {Liivam{\"a}gi} L.~J.,  {Suhhonenko} I.,   {Einasto} M.,  2019,
  \mn@doi [\aap] {10.1051/0004-6361/201936054}, \href
  {https://ui.adsabs.harvard.edu/abs/2019A&A...630A..62E} {630, A62}

\bibitem[\protect\citeauthoryear{Fonseca, Silva, Santos  \& Cooray}{Fonseca
  et~al.}{2016}]{Fonseca2016}
Fonseca J.,  Silva M.~B.,  Santos M.~G.,   Cooray A.,  2016, \mn@doi [Monthly
  Notices of the Royal Astronomical Society] {10.1093/mnras/stw2470}, 464,
  1948–1965

\bibitem[\protect\citeauthoryear{Gong, Cooray, Silva, Santos, Bock, Bradford
  \& Zemcov}{Gong et~al.}{2011}]{Gong2011}
Gong Y.,  Cooray A.,  Silva M.,  Santos M.~G.,  Bock J.,  Bradford C.~M.,
  Zemcov M.,  2011, \mn@doi [The Astrophysical Journal]
  {10.1088/0004-637x/745/1/49}, 745, 49

\bibitem[\protect\citeauthoryear{{Gong}, {Cooray}, {Silva}, {Zemcov}, {Feng},
  {Santos}, {Dore}  \& {Chen}}{{Gong} et~al.}{2017}]{2017ApJ...835..273G}
{Gong} Y.,  {Cooray} A.,  {Silva} M.~B.,  {Zemcov} M.,  {Feng} C.,  {Santos}
  M.~G.,  {Dore} O.,   {Chen} X.,  2017, \mn@doi [\apj]
  {10.3847/1538-4357/835/2/273}, \href
  {https://ui.adsabs.harvard.edu/abs/2017ApJ...835..273G} {835, 273}

\bibitem[\protect\citeauthoryear{Gong, Chen  \& Cooray}{Gong
  et~al.}{2020}]{Gong2020}
Gong Y.,  Chen X.,   Cooray A.,  2020, \mn@doi [The Astrophysical Journal]
  {10.3847/1538-4357/ab87a0}, 894, 152

\bibitem[\protect\citeauthoryear{Hamilton}{Hamilton}{1998}]{Hamilton_1998}
Hamilton A. J.~S.,  1998, in , Astrophysics and Space Science Library.
Springer Netherlands, pp 185--275, \mn@doi{10.1007/978-94-011-4960-0_17}, \url
  {https://doi.org/10.1007%2F978-94-011-4960-0_17}

\bibitem[\protect\citeauthoryear{Ho et~al.,}{Ho et~al.}{2009}]{YT2009}
Ho P. T.~P.,  et~al., 2009, \mn@doi [The Astrophysical Journal]
  {10.1088/0004-637x/694/2/1610}, 694, 1610–1618

\bibitem[\protect\citeauthoryear{Hojjati, Pogosian  \& Zhao}{Hojjati
  et~al.}{2011}]{Hojjati2011}
Hojjati A.,  Pogosian L.,   Zhao G.-B.,  2011, \mn@doi [Journal of Cosmology
  and Astroparticle Physics] {10.1088/1475-7516/2011/08/005}, 2011, 005–005

\bibitem[\protect\citeauthoryear{{Horndeski}}{{Horndeski}}{1974}]{Horndeski_1974}
{Horndeski} G.~W.,  1974, \mn@doi [International Journal of Theoretical
  Physics] {10.1007/BF01807638}, \href
  {https://ui.adsabs.harvard.edu/abs/1974IJTP...10..363H} {10, 363}

\bibitem[\protect\citeauthoryear{Ihle et~al.,}{Ihle et~al.}{2021}]{Havard_2021}
Ihle H.~T.,  et~al., 2021, COMAP Early Science: IV. Power Spectrum Methodology
  and Results, \mn@doi{10.48550/ARXIV.2111.05930}, \url
  {https://arxiv.org/abs/2111.05930}

\bibitem[\protect\citeauthoryear{{Ivezi{\'c}} et~al.,}{{Ivezi{\'c}}
  et~al.}{2019}]{2019ApJ...873..111I}
{Ivezi{\'c}} {\v Z}.,  et~al., 2019, \mn@doi [\apj] {10.3847/1538-4357/ab042c},
  \href {http://adsabs.harvard.edu/abs/2019ApJ...873..111I} {873, 111}

\bibitem[\protect\citeauthoryear{Karkare \& Bird}{Karkare \&
  Bird}{2018}]{Karkare_2018}
Karkare K.~S.,  Bird S.,  2018, \mn@doi [Physical Review D]
  {10.1103/physrevd.98.043529}, 98

\bibitem[\protect\citeauthoryear{Karkare et~al.,}{Karkare
  et~al.}{2020}]{Karkare2020}
Karkare K.~S.,  et~al., 2020, \mn@doi [Journal of Low Temperature Physics]
  {10.1007/s10909-020-02407-4}, 199, 849–857

\bibitem[\protect\citeauthoryear{{Karkare} et~al.,}{{Karkare}
  et~al.}{2022}]{Karkare_2022}
{Karkare} K.~S.,  et~al., 2022, \mn@doi [Journal of Low Temperature Physics]
  {10.1007/s10909-022-02702-2}, \href
  {https://ui.adsabs.harvard.edu/abs/2022JLTP..tmp...61K} {}

\bibitem[\protect\citeauthoryear{{Keating}, {Marrone}, {Bower}, {Leitch},
  {Carlstrom}  \& {DeBoer}}{{Keating} et~al.}{2016}]{2016ApJ...830...34K}
{Keating} G.~K.,  {Marrone} D.~P.,  {Bower} G.~C.,  {Leitch} E.,  {Carlstrom}
  J.~E.,   {DeBoer} D.~R.,  2016, \mn@doi [\apj] {10.3847/0004-637X/830/1/34},
  \href {https://ui.adsabs.harvard.edu/abs/2016ApJ...830...34K} {830, 34}

\bibitem[\protect\citeauthoryear{{Keating}, {Marrone}, {Bower}  \&
  {Keenan}}{{Keating} et~al.}{2020}]{Keating2020}
{Keating} G.~K.,  {Marrone} D.~P.,  {Bower} G.~C.,   {Keenan} R.~P.,  2020,
  \mn@doi [\apj] {10.3847/1538-4357/abb08e}, \href
  {https://ui.adsabs.harvard.edu/abs/2020ApJ...901..141K} {901, 141}

\bibitem[\protect\citeauthoryear{{Kovetz} et~al.,}{{Kovetz}
  et~al.}{2017}]{Kovetz_2017}
{Kovetz} E.~D.,  et~al., 2017, arXiv e-prints, \href
  {https://ui.adsabs.harvard.edu/abs/2017arXiv170909066K} {p. arXiv:1709.09066}

\bibitem[\protect\citeauthoryear{Kreisch \& Komatsu}{Kreisch \&
  Komatsu}{2018}]{Kreisch_2018}
Kreisch C.,  Komatsu E.,  2018, \mn@doi [Journal of Cosmology and Astroparticle
  Physics] {10.1088/1475-7516/2018/12/030}, 2018, 030–030

\bibitem[\protect\citeauthoryear{{Laureijs} et~al.,}{{Laureijs}
  et~al.}{2011}]{2011arXiv1110.3193L}
{Laureijs} R.,  et~al., 2011, arXiv e-prints, \href
  {https://ui.adsabs.harvard.edu/abs/2011arXiv1110.3193L} {p. arXiv:1110.3193}

\bibitem[\protect\citeauthoryear{Li \& Koyama}{Li \&
  Koyama}{2019}]{doi:10.1142/11090}
Li B.,  Koyama K.,  2019, Modified Gravity.
WORLD SCIENTIFIC (\mn@eprint {}
  {https://www.worldscientific.com/doi/pdf/10.1142/11090}),
  \mn@doi{10.1142/11090}, \url
  {https://www.worldscientific.com/doi/abs/10.1142/11090}

\bibitem[\protect\citeauthoryear{Li, Wechsler, Devaraj  \& Church}{Li
  et~al.}{2016}]{Li2016}
Li T.~Y.,  Wechsler R.~H.,  Devaraj K.,   Church S.~E.,  2016, \mn@doi [The
  Astrophysical Journal] {10.3847/0004-637x/817/2/169}, 817, 169

\bibitem[\protect\citeauthoryear{Lidz \& Taylor}{Lidz \&
  Taylor}{2016}]{Lidz_2016}
Lidz A.,  Taylor J.,  2016, \mn@doi [The Astrophysical Journal]
  {10.3847/0004-637x/825/2/143}, 825, 143

\bibitem[\protect\citeauthoryear{Mancini et~al.,}{Mancini
  et~al.}{2019}]{Spurio_Mancini_2019}
Mancini A.~S.,  et~al., 2019, \mn@doi [Monthly Notices of the Royal
  Astronomical Society] {10.1093/mnras/stz2581}, 490, 2155

\bibitem[\protect\citeauthoryear{{Mantz}, {Allen}, {Ebeling}  \&
  {Rapetti}}{{Mantz} et~al.}{2008}]{2008MNRAS.387.1179M}
{Mantz} A.,  {Allen} S.~W.,  {Ebeling} H.,   {Rapetti} D.,  2008, \mn@doi
  [\mnras] {10.1111/j.1365-2966.2008.13311.x}, \href
  {https://ui.adsabs.harvard.edu/abs/2008MNRAS.387.1179M} {387, 1179}

\bibitem[\protect\citeauthoryear{McQuinn, Zahn, Zaldarriaga, Hernquist  \&
  Furlanetto}{McQuinn et~al.}{2006}]{McQuinn_2006}
McQuinn M.,  Zahn O.,  Zaldarriaga M.,  Hernquist L.,   Furlanetto S.~R.,
  2006, \mn@doi [The Astrophysical Journal] {10.1086/505167}, 653, 815–834

\bibitem[\protect\citeauthoryear{Moradinezhad~Dizgah, Keating  \&
  Fialkov}{Moradinezhad~Dizgah et~al.}{2019}]{Dizgah2019}
Moradinezhad~Dizgah A.,  Keating G.~K.,   Fialkov A.,  2019, \mn@doi [The
  Astrophysical Journal] {10.3847/2041-8213/aaf813}, 870, L4

\bibitem[\protect\citeauthoryear{{Nadolski} et~al.,}{{Nadolski}
  et~al.}{2020}]{nadolski2019}
{Nadolski} A.,  et~al., 2020, \mn@doi [\ao] {10.1364/AO.383921}, \href
  {https://ui.adsabs.harvard.edu/abs/2020ApOpt..59.3285N} {59, 3285}

\bibitem[\protect\citeauthoryear{{Noller}}{{Noller}}{2020}]{2020PhRvD.101f3524N}
{Noller} J.,  2020, \mn@doi [\prd] {10.1103/PhysRevD.101.063524}, \href
  {https://ui.adsabs.harvard.edu/abs/2020PhRvD.101f3524N} {101, 063524}

\bibitem[\protect\citeauthoryear{Noller \& Nicola}{Noller \&
  Nicola}{2019}]{Noller_2019}
Noller J.,  Nicola A.,  2019, \mn@doi [Physical Review D]
  {10.1103/physrevd.99.103502}, 99

\bibitem[\protect\citeauthoryear{Padmanabhan}{Padmanabhan}{2017}]{Padmanahhan2017}
Padmanabhan H.,  2017, \mn@doi [Monthly Notices of the Royal Astronomical
  Society] {10.1093/mnras/stx3250}, 475, 1477–1484

\bibitem[\protect\citeauthoryear{Padmanabhan, Breysse, Lidz  \&
  Switzer}{Padmanabhan et~al.}{2021}]{padmanabhan2021intensity}
Padmanabhan H.,  Breysse P.,  Lidz A.,   Switzer E.~R.,  2021, Intensity
  mapping from the sky: synergizing the joint potential of [OIII] and [CII]
  surveys at reionization (\mn@eprint {arXiv} {2105.12148})

\bibitem[\protect\citeauthoryear{Paine}{Paine}{2022}]{paine_2022}
Paine S.,  2022, The am atmospheric model (12.0).

\bibitem[\protect\citeauthoryear{Peirone, Koyama, Pogosian, Raveri  \&
  Silvestri}{Peirone et~al.}{2018}]{P2018}
Peirone S.,  Koyama K.,  Pogosian L.,  Raveri M.,   Silvestri A.,  2018,
  \mn@doi [Physical Review D] {10.1103/physrevd.97.043519}, 97

\bibitem[\protect\citeauthoryear{{Perlmutter} et~al.,}{{Perlmutter}
  et~al.}{1999}]{Perlmutter_1999}
{Perlmutter} S.,  et~al., 1999, \mn@doi [\apj] {10.1086/307221}, \href
  {https://ui.adsabs.harvard.edu/abs/1999ApJ...517..565P} {517, 565}

\bibitem[\protect\citeauthoryear{Pullen, Dor{\'{e} }  \& Bock}{Pullen
  et~al.}{2014}]{Pullen_2014}
Pullen A.~R.,  Dor{\'{e} } O.,   Bock J.,  2014, \mn@doi [The Astrophysical
  Journal] {10.1088/0004-637x/786/2/111}, 786, 111

\bibitem[\protect\citeauthoryear{Rapetti, Allen  \& Mantz}{Rapetti
  et~al.}{2008}]{Rap2008}
Rapetti D.,  Allen S.~W.,   Mantz A.,  2008, \mn@doi [Monthly Notices of the
  Royal Astronomical Society] {10.1111/j.1365-2966.2008.13460.x}, 388,
  1265–1278

\bibitem[\protect\citeauthoryear{{Riess} et~al.,}{{Riess}
  et~al.}{1998}]{Riess_1998}
{Riess} A.~G.,  et~al., 1998, \mn@doi [\aj] {10.1086/300499}, \href
  {https://ui.adsabs.harvard.edu/abs/1998AJ....116.1009R} {116, 1009}

\bibitem[\protect\citeauthoryear{Righi, Hernández-Monteagudo  \&
  Sunyaev}{Righi et~al.}{2008}]{Righi_2008}
Righi M.,  Hernández-Monteagudo C.,   Sunyaev R.~A.,  2008, \mn@doi [Astronomy
  & Astrophysics] {10.1051/0004-6361:200810199}, 489, 489–504

\bibitem[\protect\citeauthoryear{Samushia, Percival  \& Raccanelli}{Samushia
  et~al.}{2012}]{Samushia_2012}
Samushia L.,  Percival W.~J.,   Raccanelli A.,  2012, \mn@doi [Monthly Notices
  of the Royal Astronomical Society] {10.1111/j.1365-2966.2011.20169.x}, 420,
  2102

\bibitem[\protect\citeauthoryear{Schaan \& White}{Schaan \&
  White}{2021}]{Schaan_2021}
Schaan E.,  White M.,  2021, \mn@doi [Journal of Cosmology and Astroparticle
  Physics] {10.1088/1475-7516/2021/05/067}, 2021, 067

\bibitem[\protect\citeauthoryear{{Schmidt} et~al.,}{{Schmidt}
  et~al.}{1998}]{Schmidt_1998}
{Schmidt} B.~P.,  et~al., 1998, \mn@doi [\apj] {10.1086/306308}, \href
  {https://ui.adsabs.harvard.edu/abs/1998ApJ...507...46S} {507, 46}

\bibitem[\protect\citeauthoryear{{Sheth}, {Mo}  \& {Tormen}}{{Sheth}
  et~al.}{2001}]{2001MNRAS.323....1S}
{Sheth} R.~K.,  {Mo} H.~J.,   {Tormen} G.,  2001, \mn@doi [\mnras]
  {10.1046/j.1365-8711.2001.04006.x}, \href
  {https://ui.adsabs.harvard.edu/abs/2001MNRAS.323....1S} {323, 1}

\bibitem[\protect\citeauthoryear{{Shirokoff} et~al.,}{{Shirokoff}
  et~al.}{2012}]{shirokoff2012}
{Shirokoff} E.,  et~al., 2012, in {Holland} W.~S.,  {Zmuidzinas} J.,  eds,
  Society of Photo-Optical Instrumentation Engineers (SPIE) Conference Series
  Vol. 8452, Millimeter, Submillimeter, and Far-Infrared Detectors and
  Instrumentation for Astronomy VI. p. 84520R (\mn@eprint {arXiv} {1211.1652}),
  \mn@doi{10.1117/12.927070}

\bibitem[\protect\citeauthoryear{Silva, Santos, Gong, Cooray  \& Bock}{Silva
  et~al.}{2013}]{Silva2013}
Silva M.~B.,  Santos M.~G.,  Gong Y.,  Cooray A.,   Bock J.,  2013, \mn@doi
  [The Astrophysical Journal] {10.1088/0004-637x/763/2/132}, 763, 132

\bibitem[\protect\citeauthoryear{Silva, santos, Cooray  \& Gong}{Silva
  et~al.}{2015}]{silva2015prospects}
Silva M.~B.,  santos M.~G.,  Cooray A.,   Gong Y.,  2015, Prospects for
  detecting CII emission during the Epoch of Reionization (\mn@eprint {arXiv}
  {1410.4808})

\bibitem[\protect\citeauthoryear{Silva, Zaroubi, Kooistra  \& Cooray}{Silva
  et~al.}{2017}]{silva2017tomographic}
Silva M.~B.,  Zaroubi S.,  Kooistra R.,   Cooray A.,  2017, Tomographic
  Intensity Mapping versus Galaxy Surveys: Observing the Universe in H-alpha
  emission with new generation instruments (\mn@eprint {arXiv} {1711.09902})

\bibitem[\protect\citeauthoryear{{Smith} et~al.,}{{Smith}
  et~al.}{2003}]{2003MNRAS.341.1311S}
{Smith} R.~E.,  et~al., 2003, \mn@doi [\mnras]
  {10.1046/j.1365-8711.2003.06503.x}, \href
  {https://ui.adsabs.harvard.edu/abs/2003MNRAS.341.1311S} {341, 1311}

\bibitem[\protect\citeauthoryear{{Spergel} et~al.,}{{Spergel}
  et~al.}{2015}]{2015arXiv150303757S}
{Spergel} D.,  et~al., 2015, arXiv e-prints, \href
  {https://ui.adsabs.harvard.edu/abs/2015arXiv150303757S} {p. arXiv:1503.03757}

\bibitem[\protect\citeauthoryear{{Sun} et~al.,}{{Sun} et~al.}{2018}]{sun2018}
{Sun} G.,  et~al., 2018, \mn@doi [\apj] {10.3847/1538-4357/aab3e3}, \href
  {https://ui.adsabs.harvard.edu/abs/2018ApJ...856..107S} {856, 107}

\bibitem[\protect\citeauthoryear{Taruya, Nishimichi  \& Saito}{Taruya
  et~al.}{2010}]{Taruya_2010}
Taruya A.,  Nishimichi T.,   Saito S.,  2010, \mn@doi [Physical Review D]
  {10.1103/physrevd.82.063522}, 82

\bibitem[\protect\citeauthoryear{Troxel et~al.,}{Troxel
  et~al.}{2018}]{Troxel2018}
Troxel M.,  et~al., 2018, \mn@doi [Physical Review D]
  {10.1103/physrevd.98.043528}, 98

\bibitem[\protect\citeauthoryear{Wang}{Wang}{2006}]{Wang_2006}
Wang Y.,  2006, \mn@doi [The Astrophysical Journal] {10.1086/505384}, 647,
  1–7

\bibitem[\protect\citeauthoryear{Wu \& Zhang}{Wu \&
  Zhang}{2021}]{wu2021prospects}
Wu P.-J.,  Zhang X.,  2021, Prospects for measuring dark energy with 21 cm
  intensity mapping experiments (\mn@eprint {arXiv} {2108.03552})

\bibitem[\protect\citeauthoryear{Yue, Ferrara, Pallottini, Gallerani  \&
  Vallini}{Yue et~al.}{2015}]{Yue2015}
Yue B.,  Ferrara A.,  Pallottini A.,  Gallerani S.,   Vallini L.,  2015,
  \mn@doi [Monthly Notices of the Royal Astronomical Society]
  {10.1093/mnras/stv933}, 450, 3829–3839

\bibitem[\protect\citeauthoryear{Zumalacárregui, Bellini, Sawicki, Lesgourgues
   \& Ferreira}{Zumalacárregui et~al.}{2017}]{Zumalacarregui_2017}
Zumalacárregui M.,  Bellini E.,  Sawicki I.,  Lesgourgues J.,   Ferreira
  P.~G.,  2017, \mn@doi [Journal of Cosmology and Astroparticle Physics]
  {10.1088/1475-7516/2017/08/019}, 2017, 019–019

\makeatother
\end{thebibliography}

%\begin{thebibliography}{99}
%\end{thebibliography}

%%%%%%%%%%%%%%%%%%%%%%%%%%%%%%%%%%%%%%%%%%%%%%%%%%

%%%%%%%%%%%%%%%%% APPENDICES %%%%%%%%%%%%%%%%%%%%%

\appendix

\section{Estimating Noise Power}
\label{app:noise_power}

Noise power can be expressed in terms of the Noise Equivalent Temperature (NET) or the Noise Equivalent Flux Density (NEFD). For consistency with our parameterization of the line luminosities in terms of line temperature ($\mu$K), we choose to work in NET. We begin by assuming a dual polarization instrument and calculate the noise equivalent power (NEP) from the incident photon load $Q$ for each detector: 
\begin{equation}
\label{eqn: NEP} 
    \mathrm{NEP}_{\mathrm{ph}} = 2h\nu Q + \frac{1}{N_{\mathrm{modes}}} \frac{2Q^2}{\Delta \nu},
\end{equation}
where $\nu$ is the detector center frequency and $\Delta \nu$ is the bandwidth. $Q$ is the sum of power arriving at the detector from the atmosphere and emission from the telescope:
\begin{equation}
\label{eqn: Q} 
Q_{\rm tot} = Q_{\rm atm} + Q_{\rm tel}.
\end{equation}
A detector observing a load of temperature $T$ with optical efficiency $\eta$ sees photon power $Q \approx 2 \eta k T \Delta \nu$ (for $h\nu \ll kT$). We use the \textit{am} atmospheric modeling software to calculate the typical atmospheric temperature at each frequency for the South Pole winter \citep{paine_2022}. The telescope emission is assumed to be at the ambient South Pole temperature $\sim 250$ K with an emission $\epsilon = 0.01$, as measured for the South Pole Telescope. Finally, we assume that each detector has an NEP of  $\sim 10^{-18}$ W/$\sqrt{\mathrm{Hz}}$, which is added in quadrature to the incident photon NEP.

After converting the NEP to a white noise level $\sigma_{\mathrm{rms}} (\approx 481$ $\mu$K $\cdot \sqrt{s}$), the noise power spectrum for a given integration time per pixel $t_{\mathrm{pix}}$ is
\begin{equation}
\label{eqn: Pn} 
    P_{\mathrm{N}} = V_{\mathrm{vox}} \frac{\sigma_{\mathrm{rms}}^2}{t_{\mathrm{pix}}},
\end{equation}
where the voxel volume is
\begin{equation}
\label{eqn: V_vox} 
    V_{\mathrm{vox}} = r(z)^2 \frac{\lambda (1+z)^2}{H(z)} \Omega_{\mathrm{pix}} \Delta \nu.
\end{equation}
Here $r(z)$ is the comoving radial distance, $\lambda$ is the wavelength, $H(z)$ is the Hubble parameter, and $\Omega_{\mathrm{pix}}$ is the pixel size.

%%%%%%%%%%%%%%%%%%%%%%%%%%%%%%%%%%%%%%%%%%%%%%%%%%

% Don't change these lines
\bsp	% typesetting comment
\label{lastpage}
\end{document}